\documentclass[12pt]{article}

\usepackage[utf8]{inputenc}
\usepackage{charter}
\usepackage{comment}
\usepackage{fullpage}
\usepackage{hyperref}
\usepackage{graphicx}
\usepackage{deluxetable}
\usepackage{lipsum}
\usepackage[sort&compress,numbers]{natbib}
\usepackage{hyperref}
\usepackage{amsmath, amssymb}
\hypersetup{hidelinks}
\usepackage[total={6.5in,9in},top=1in,headsep=0.1in,headheight=1in]{geometry}

\makeatletter
\def\@to{to}
\makeatother

\def\aapr{\ref@jnl{A\&A~Rev.}}          

\newcommand{\snia}{Type Ia SN}

\newcommand{\h}{\ensuremath{\mathrm{H}_0}}
\newcommand{\lcdm}{\ensuremath{\Lambda}CDM}
\newcommand{\mpckms}{\ensuremath{\mathrm{km}/\mathrm{s}/\mathrm{Mpc}}}

\newcommand\snowmass{
\begin{center}
  \rule[-0.2in]{\hsize}{0.01in}\\
  \rule{\hsize}{0.01in}\\
  \vskip 0.1in
  Submitted to the Proceedings of the US Community Study\\ 
  on the Future of Particle Physics (Snowmass 2021)\\
  \rule{\hsize}{0.01in}\\
  \rule[+0.2in]{\hsize}{0.01in}\\[-2em]
\end{center}
}

\usepackage[firstpage=true]{background}
\backgroundsetup{contents={\parbox{6.5in}{\snowmass}}, scale=1,placement=top,opacity=1,color=black,position={3.25in,1.2in}}

\usepackage{fancyhdr}
\fancypagestyle{plain}{%
  \fancyhf{}%
  \fancyhead[C]{}
  \fancyfoot[C]{\thepage}
}

\fancypagestyle{empty}{%
  \fancyhf{}%
  \fancyhead[C]{{\it Snowmass2021 Cosmic Frontier White Paper: Enabling Flagship Dark Energy Experiments}}
  \fancyfoot[C]{\thepage}
}
\title{Snowmass2021 Cosmic Frontier White Paper: Enabling Flagship Dark Energy Experiments to Reach their Full Potential}
\date{}

\usepackage{authblk}

\author[1]{Jonathan A.~Blazek}
\author[2]{Doug Clowe}
\author[3]{Thomas E.~Collett}
\author[4]{Ian P.~Dell’Antonio}
\author[5]{Mark Dickinson}
\author[6,7]{Llu\'{i}s Galbany}
\author[8]{Eric~Gawiser}
\author[9]{Katrin Heitmann}
\author[10]{Ren\'ee Hlo\v{z}ek}
\author[11]{Mustapha Ishak}
\author[8]{Saurabh~W.~Jha}
\author[12]{Alex G.~Kim}
\author[13]{C. Danielle Leonard}
\author[14]{Anja von der Linden}
\author[15,16]{Michelle Lochner}
\author[17]{Rachel~Mandelbaum}
\author[18]{Peter Melchior}
\author[19]{Joel~Meyers}
\author[20]{Jeffrey A.~Newman}
\author[12,21]{Peter Nugent}
\author[12,22]{Saul Perlmutter}
\author[20,23]{Daniel J.~Perrefort}
\author[24]{Javier~S\'{a}nchez}
\author[25]{Samuel J.~Schmidt}
\author[17]{Sukhdeep Singh}
\author[26]{Mark Sullivan}
\author[27]{Aprajita Verma}
\author[12]{Rongpu Zhou}

\affil[1]{Department of Physics, Northeastern University, Boston, MA, 02115, USA}

\affil[2]{Ohio University, Department of Physics and Astronomy, Clippinger Labs 251B, Athens, OH 45701, USA}

\affil[3]{Institute of Cosmology and Gravitation, University of Portsmouth, Portsmouth PO1 3FX, UK}

\affil[4]{Physics Department, Brown University, Providence, RI, 02912, USA}

\affil[5]{NSF's National Optical-Infrared Astronomy Research Laboratory, Tucson, AZ, USA}

\affil[6]{Institute of Space Sciences (ICE, CSIC), Campus UAB, Carrer de Can Magrans, s/n, E-08193 Barcelona, Spain}
\affil[7]{Institut d’Estudis Espacials de Catalunya (IEEC), E-08034 Barcelona, Spain}

\affil[8]{Department of Physics and Astronomy, Rutgers, the State University of New Jersey, Piscataway, NJ 08854, USA}

\affil[9]{Argonne National Laboratory, 9700 S. Cass Ave, Lemont, IL 60439, USA}

\affil[10]{David A. Dunlap Department of Astronomy and Astrophysics \&  Dunlap Institute for Astronomy and Astrophysics, University of Toronto, 50 St. George Street, M5S 3H4, Canada} 

\affil[11]{Department of Physics, The University of Texas at Dallas, Richardson, Texas 75080, USA}

\affil[8]{Department of Physics and Astronomy, Rutgers, the State University of New Jersey, Piscataway, NJ 08854, USA}

\affil[12]{Lawrence Berkeley National Laboratory, 1 Cyclotron Road, Berkeley, CA 94720, USA}

\affil[13]{School of Mathematics, Statistics and Physics, Herschel Building, Newcastle University, Newcastle-upon-Tyne, NE1 7RU, United Kingdom}
\affil[14]{Department of Physics and Astronomy, Stony Brook University, Stony Brook University, Stony Brook, NY 11794, USA}

\affil[15]{Department of Physics and Astronomy, University of the Western Cape, Bellville, Cape Town, 7535, South Africa}
\affil[16]{South African Radio Astronomy Observatory (SARAO), The Park, Park Road, Pinelands, Cape Town 7405, South Africa}

\affil[17]{McWilliams Center for Cosmology, Department of Physics, Carnegie Mellon University, Pittsburgh, PA 15213, USA}

\affil[18]{Department of Astrophysical Sciences, Princeton University, Princeton, NJ 08540, USA}

\affil[19]{Department of Physics, Southern Methodist University, Dallas, TX 75275, USA}

\affil[20]{Department of Physics and Astronomy and PITT PACC, University of Pittsburgh, Pittsburgh, PA 15260, USA}

\affil[21]{Department of Astronomy, University of California at Berkeley, New Campbell Hall,
Berkeley, CA 94720, USA}

\affil[22]{Department of Physics, University of California Berkeley, 366 LeConte Hall MC 7300, Berkeley, CA, 94720, USA}

\affil[23]{Center for Research Computing, University of Pittsburgh, Pittsburgh, PA 15260, USA}

\affil[24]{Fermi National Accelerator Laboratory, PO Box 500, Batavia, IL 60510, USA}

\affil[25]{Department of Physics, University of California, Davis, One Shields Avenue, Davis, CA 95616, USA}

\affil[26]{School of Physics and Astronomy, Univeristy of Southampton, Southampton, SO17 1BJ, UK}

\affil[27]{Sub-department of Astrophysics, University of Oxford, Denys Wilkinson Building, Keble Road, Oxford OX1 3RH, UK}

\begin{document}

\maketitle

\begin{abstract}

A new generation of powerful dark energy experiments will open new vistas for cosmology in the next decade. However, these projects cannot reach their utmost potential without data from other telescopes. This white paper focuses in particular on the compelling benefits of ground-based spectroscopic and photometric observations to complement the Vera C. Rubin Observatory, as well as smaller programs in aid of a DESI-2 experiment and CMB-S4. These additional data sets will both improve dark energy constraints from these flagship projects beyond what would possible on their own and open completely new windows into fundamental physics. 
For example, additional photometry and single-object spectroscopy will provide necessary follow-up information for supernova and strong lensing cosmology, while highly-multiplexed spectroscopy both from smaller facilities over wide fields and from larger facilities over narrower regions of sky will yield more accurate photometric redshift estimates for weak lensing and galaxy clustering measurements from the Rubin Observatory, provide critical spectroscopic host galaxy redshifts for supernova Hubble diagrams, provide improved understanding of limiting astrophysical systematic effects, and enable new measurements that probe the nature of gravity. A common thread is that access to complementary data from a range of telescopes/instruments would have a substantial impact on the rate of advance of dark energy science in the coming years.

\end{abstract}




\section{Introduction}

The newest generation of ``Stage IV'' Dark Energy experiments such as the Vera C. Rubin Observatory Legacy Survey of Space and Time (LSST) and the Dark Energy Spectroscopic Instrument (DESI) will play a major role in improving our knowledge of cosmology in the coming years. However, additional data from other ground-based facilities can enable these facilities to constrain models of cosmic acceleration more strongly and reach their utmost potential.  

The gains from additional data can arise in several ways.  Strong lensing and supernova cosmology constraints from LSST can be greatly strengthened by making use of additional telescopes and instruments that will enable detailed follow-up of individual objects. These facilities can provide \textbf{single-object imaging} either in more bands, with higher spatial resolution, or with higher cadence than the LSST observations provide, and can provide \textbf{spectroscopic measurements} for rare objects, which are difficult to target efficiently for multi-object spectroscopy.  
Spectroscopy of individual strong lens systems will constrain lens properties and measure redshifts for both the lenses and background objects; spectroscopy can also classify individual supernovae in ``live'' follow up observations.  

Other opportunities to enhance cosmological measurements from LSST  would be enabled by access to {\bf deep ($i \sim 25$), highly-multiplexed} optical and near-infrared multi-object spectroscopy on 8--40~m telescopes. Such capabilities would enable improved photo-$z$ estimates and reduce systematic uncertainties \cite{descsrd}.  Every major cosmological probe based upon LSST data would benefit.  
As an example, photometric redshift (photo-$z$) estimates depend on deep spectroscopic datasets for optimization of algorithms; every major cosmological probe will rely on photo-$z$'s directly or indirectly.  However, focused spectroscopic campaigns can also improve weak lensing cosmology by constraining the intrinsic alignments between the orientations of galaxies; contribute to galaxy cluster studies by measuring kinematics of galaxies in clusters and testing photo-$z$ performance in regions of high density; and aid strong lensing measurements by determining the distribution of foreground mass along the lines of sight to systems.  

A third area of gains to LSST would come from access to {\bf wide-field ($>20$ deg$^2$ total survey area), highly-multiplexed} optical and near-infrared multi-object spectroscopy (MOS) on 4--15~m telescopes.  LSST photo-$z$'s can be calibrated with high precision via cross-correlations against large-area spectroscopic samples.  Spectroscopy with wide-field instruments can also constrain galaxy intrinsic alignments using cross-correlations with shallower spectroscopic samples; aid strong lensing measurements by providing a first characterization of redshifts of candidate strong lens systems; provide tests of modified gravity via a variety of cross-correlation techniques; and contribute to supernova cosmology by providing redshifts for host galaxies of supernovae that lack direct spectroscopy. 

However, the Vera Rubin Observatory will not be the only beneficiary from additional data obtained with other facilities.  As an example, future surveys with the DESI instrument can more efficiently target objects of interest if deep, multiband imaging (such as that which LSST will produce) is available in regions north of the coverage of the main LSST ``Wide, Fast, Deep'' survey.  Additionally, cosmology with galaxy clusters from CMB-S4 will benefit from deep multiplexed spectroscopy, overlapping with the needs for LSST cluster science.
Additional discussions of the benefits to cosmological studies from combining datasets from multiple surveys are provided in two other Snowmass white papers \cite{Snowmass2021:JointProbes, Snowmass2013:Transient}.

In the remainder of this white paper, we describe the benefits that additional data can bring to upcoming Dark Energy experiments, considering the gains for each science case separately.  We concentrate this discussion on upcoming experiments that are primarily focused on constraining the nature of Dark Energy, but note that other cosmology experiments should also benefit from access to the additional capabilities we outline here.


\section{Enabling Measurements of the Supernova Distance -- Redshift Relation with LSST}

\label{sec:singlesn}

\subsection{Photometric follow-up} 

For supernova cosmology, we require measurements of supernova light curves (brightness as a function of time) in multiple bands with dense enough time sampling to accurately determine the peak brightness of the supernova and the rate at which it declines.  For some supernovae observed by LSST, the standard ``Wide, Fast, Deep'' cadence will not be optimized for these measurements, and additional photometry from other facilities can enhance supernova cosmology in the coming decade.

There are two major ways in which additional photometry in the LSST fields can benefit supernova cosmology. First, the cadence of the LSST will not be optimal for all transients of interest. 
Many LSST science cases are focused on high-precision photometry and astrometry spanning long periods of time for transients which are brighter than cosmological supernovae (e.g., variable stars, high-proper motion stars, etc.), while on the fainter end the most common transients of interest have light curves that evolve only slowly (e.g., high-redshift Type Ia supernovae for cosmology). 
The need to serve these science cases results in the standard LSST cadence being less optimal for some classes of supernovae.  Higher-cadence observations can complement LSST by providing higher-quality light curves for nearby supernovae, improving cosmological measurements. 

Additionally, it is easier to obtain follow-up data (e.g., spectroscopy) for more nearby, brighter objects, as then smaller and less expensive telescopes can be put to use.
Obtaining a timely spectrum is a key need for much supernova science (as described below).   
However, much of the very nearby universe is inaccessible to the Rubin Observatory as bright objects will saturate its detectors. As a result a high-cadence, shallow survey is an optimal complement to Rubin and the LSST for nearby supernova science such as the measurement of peculiar velocities. 

ZTF, while focused in the north, is one such survey\citep{ztfnat}. In the south, there is the La Silla Schmidt Southern Survey \citep[LS4;][]{LOI-LS4}. LS4 is a 5-year public, wide-field, optical survey using the upgraded 20 square degree QUEST Camera on the ESO Schmidt Telescope at the La Silla Observatory in Chile.  It uses LBNL deep-depletion CCDs to maximize the sensitivity in the optical up to 1\,micron. As outlined above, it complements the LSST by providing a higher cadence over several thousand square degrees of sky in the $g-$ and $z$-bands, allowing a more accurate characterization of brighter and faster evolving transients down to 21$^{st}$ magnitude. It also opens up a new phase-space for discovery when coupled with the LSST by probing the sky between 12--16$^{th}$ magnitude -- a region where the Rubin Observatory saturates. As such, it will greatly improve both the identification and characterization of nearby Type Ia supernova for several of the cosmology studies outlined below.

\subsection{Single-object spectroscopic observations of active SNe}
As described in \autoref{sec:widesn}, the largest sets of redshift measurements for LSST supernovae should come from targeting SNe and their host galaxies via wide-field multi-object spectroscopy, either using a small fraction of fibers in surveys that span the LSST footprint or via targeted surveys in the LSST Deep Drilling Fields (DDFs).  However, higher signal-to-noise observations will be needed for a significant sample of low-$z$ SNe to characterize the range of intrinsic spectra; and to compare this to the high-$z$ LSST \snia\ ($z \sim 0.5 - 1.2$) will require targeted long-exposure follow-up while the high-$z$ SNe are still bright.

\textbf{Typing to eliminate non-SNe~Ia contamination and to validate photometric classification at high (and low) redshift:}  Much work has gone into techniques for determining supernova types and redshifts from photometry alone, but this is still a difficult technique when applied over a large redshift range due to changing filter coverage and unknown drifts in the spectroscopic properties of the SNe.  Current photometric identification techniques require spectroscopically classified training samples, ideally over the full range of redshifts under consideration.  Fully characterizing some classes of objects from LSST at the faintest magnitudes \citep[e.g., SLSN and Type II SN, see][]{deJaeger2017, inserra2018} will require spectroscopic observations on 25--40~m class telescopes. Furthermore, measuring the redshifts via spectra of ``live'' supernovae eliminates type uncertainty, yielding high-purity samples with limited systematics. Ultimately, detailed spectrophotometric studies of low-$z$ \snia~will be the best source of high-signal-to-noise information about the variations over a large range of \snia~subtypes, making spectroscopy on smaller telescopes valuable as well. An example of such a study can be found in Ref. \cite{Saunders2018}.

\textbf{Systematic error mitigation:}
Detailed understanding of a variety of potential sources of systematic error will be crucial for supernova science with LSST, including evolution of the \snia\  population with redshift; errors in the specification of the spectral energy distribution (SED) in different redshift/rest-frame wavelength ranges that can affect training of light curve models; and mis-classifying fine sub-types (``twins'') of SNe Ia. 
Finally, we do not know if the current catalog of spectral properties is exhaustive, and, if not, the mismatch between assumed spectral model templates and actual spectra will introduce systematic shifts in  cosmological distance measurements. While deep follow-up of a variety of LSST transients will be needed to understand their physics and evolution, requirements for the use of SNe Ia as cosmic probes of distance are particularly stringent, and hence detailed spectral studies are needed.

Very high data quality is needed if spectra of ``live" supernovae will be used to detect the subtle differences between sub-types of SNe Ia, to study evolutionary population drift, and to match highly  similar supernovae (``twins'' as described in \cite{Fakhouri2015}) to reduce the dispersion of measured distances. Such purposes require high signal-to-noise observations to track subtle spectroscopic features and spectrophotometric data to allow a clean subtraction of the host galaxy light.  For systematics constraints on measurements of high-redshift \snia\ to be commensurate with the precise statistical uncertainties resulting from large numbers of LSST-discovered SNe, new instrumentation for ground-based telescopes will be needed (e.g., IFU or high-throughput spectrographs now in the planning stages for 2--4~m telescopes and IFU reformatters on existing spectrographs, with or without AO, on 8--40~m telescopes). The development of space-based observational avenues such as coordinated observations with Euclid, WFIRST or other space missions now planning spectrophotometric instrumentation could also be helpful.


\subsection{Multi-object spectroscopic observations of supernovae and their host galaxies}

\label{sec:widesn}

SNe Ia provide a mature and well-exploited probe of the accelerating universe (e.g., \cite{2018arXiv181102374D}), and their use as standardizable candles is an immediate route to measuring the equation of state of dark energy. LSST, for example, could assemble around 100,000 SNe Ia to $z=1$, giving unprecedented insight into the expansion history of the universe. A major systematic uncertainty will be the photometric classification and redshift measurement of the supernova detections. Wide-field spectroscopy can exploit the fact that wherever a follow-up facility points in the extragalactic sky, there will be known time-variable sources, including both recently discovered transients and older, now-faded events.

Spectroscopy serves two main goals over both the typical LSST Wide, Fast, Deep (WFD) footprint and the more frequently-observed Deep Drilling Fields (DDFs).  
The first is the classification of live SNe and the construction of optimized training samples for photometric classifiers to assemble the next generation of SN Ia cosmological samples. 
Even the most advanced machine-learning classification techniques are fundamentally dependent on large, homogeneous and representative training sets \citep{2016ApJS..225...31L}. 
%
The second goal is to obtain spectroscopic redshifts (spec-$z$'s) for host galaxies of SNe that have faded away. While conventional SN~Ia cosmology analyses rely on spectroscopic follow-up of all the SNe, new analyses \citep[e.g.][]{hlozek2012,campbell2013,jones2018} show the possibility of taking advantage of even larger samples of SNe after obtaining spec-$z$'s of their host galaxies. Targeted campaigns of multi-object spectroscopy of SNe and their hosts in LSST DDFs, combined with the assignment of fibers to such objects in larger surveys covering the wide LSST footprint, can provide the redshifts needed for most LSST SN Ia cosmology studies. 
More detailed investigation of a smaller set of SNe will remain valuable, however, as described in \autoref{sec:singlesn}.


\subsubsection{Supernovae and their hosts in LSST Deep Drilling Fields (DDFs)}

\label{DDF_hosts}


The best-characterized and deepest LSST SN samples will come from the higher-quality light curves provided by more frequent and deeper photometry  of the DDFs.  Because of the comparatively small area covered by these fields ($\sim50$ deg$^2$ total) and the low surface density of SNe, it is feasible to obtain long exposures on all DDF live SNe or their hosts using  wide-field spectrographs (e.g., DESI or Subaru/PFS in the North and 4MOST in the South); c.f.~\cite{kavli}. This approach has proven very successful in the OzDES survey \cite{ozdes}, and will be employed by the 4MOST/TiDES survey for LSST SNe \cite{tides}. 
In addition to the cosmological measurements enabled by DDF SNe and host redshifts, the resulting dataset should be an extremely valuable source of high signal-to-noise templates for light curves to be used in cosmology for SNe that have LSST data but no spectroscopy, and will also provide a wealth of training data for photometric classification of transients.

Such surveys of supernova hosts can be performed efficiently at a variety of current (e.g., DESI) or proposed facilities.  Table \ref{table:supernova_times} describes the total amount of time it would take with different instruments and telescopes to perform annual spectroscopy of 100 new	$r<24$	galaxy	hosts of supernovae	per	square degree spanning five LSST deep drilling fields, each 10 square degrees in area; characteristics of relevant facilities are summarized in \cite{Chakrabarti2022}.  We scale the exposure time for each facility to be equivalent to 8 hours per pointing on a 4m telescope, which should yield sufficient signal-to-noise to obtain secure redshifts for at least 75\% of such targets (Supernova hosts that fail to	yield redshifts could be retargeted in successive years to further improve S/N).  
Over the course of the 10-year LSST survey, such
follow-up would provide redshifts for most of the $\sim 50,000$ best-characterized LSST SNe Ia.  
Note that the time costs can be
mitigated by incorporating SN host targets
into a denser MOS observing program serving multiple science needs.  Indeed OzDES survey exposure times simultaneously  served
SN host-galaxy redshifting,
photo-z calibration, active-transient classification, and QSO monitoring.

Details of the assumptions made about how instruments may be compared in these calculations are given in \cite{specneeds}.  Results are given in ``dark-years'': one dark-year consists of 365 8-hour-long dark nights, of which one-third are lost to weather and instrumental overheads.


\begin{figure}[htbp!]
\begin{center}
\includegraphics[width=1.0\textwidth,trim=0cm 14.5cm 0.0cm 0.0cm, clip]{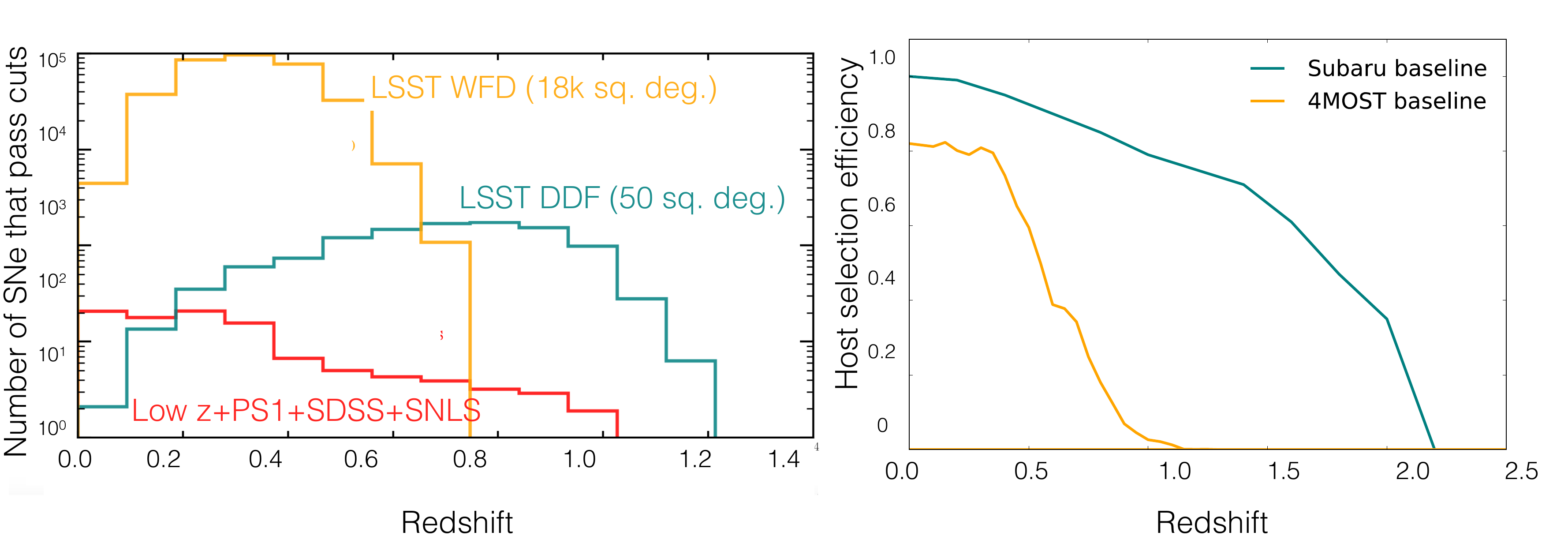}
\caption{\footnotesize \textit{Left panel:} The redshift distributions of high-quality supernovae that should be observed in the wide (WFD) and deep (DDF) regions of the LSST survey. LSST will vastly increase SN samples compared to current surveys; spec-$z$'s will greatly increase the value of the SNe for cosmological analyses. \textit{Right panel:} The fraction of SNe Ia expected to have secure redshift measurements from 4MOST/TiDES observations of $i < 22$ SN hosts in the WFD region (orange curve), as well as expectations for an $i < 22.9$ Subaru/PFS sample in the DDFs (blue curve); the existence of such data is assumed in Ref.~\cite{descsrd}, and will be vital for SN Ia cosmology with LSST.  \vspace{-0.1in}}
 \label{fig:sn_specsize}
 \end{center}
 \end{figure}

\begin{deluxetable}{lllll}
\tabletypesize{\footnotesize}
\tablecaption{Time required per epoch of SN host spectroscopy in LSST deep fields
\label{table:supernova_times}}
\tablehead{ \colhead{Instrument / Telescope} & \colhead{Collecting}  & \colhead{Field area}  & \colhead{Multiplex} & \colhead{Total time} \\
 & \colhead{Area (sq. m)} & \colhead{(sq. arcmin)} &  & \colhead{(dark-years)} }
 \startdata
        4MOST & 10.7 & 14,400 & 1,400 & 0.05 \\ 
        Mayall 4m / DESI & 11.4 & 25,500 & 5,000 & 0.03 \\
        WHT / WEAVE & 13.0 & 11,300 & 1,000 & 0.06 \\ 
        Megamapper (Magellan-like) & 28.0 & 25,416 & 20,000 & 0.01 \\ 
        Subaru / PFS & 53.0 & 4,500 & 2,400 & 0.04 \\ 
        VLT / MOONS & 58.2 & 500 & 500 & 0.29 \\ 
        Keck / DEIMOS & 76.0 & 54 & 150 & 2.04 \\ 
        Keck / FOBOS & 76.0 & 314 & 1,800 & 0.35 \\ 
        ESO SpecTel & 87.9 & 17,676 & 15,000 & 0.01 \\ 
        MSE & 97.6 & 6,359 & 3,249 & 0.01 \\ 
        GMT/MANIFEST + GMACS & 368.0 & 314 & 420 & 0.07 \\
        TMT / WFOS & 655.0 & 25 & 100 & 0.51 \\ 
        E-ELT / Mosaic Optical & 978.0 & 39 & 200 & 0.22\tablenotemark{1} \\ 
        E-ELT / MOSAIC NIR & 978.0 & 46 & 100 & 0.19
\enddata
\tablenotetext{1}{For E-ELT, observations in both the optical and near-IR settings are required to achieve the required wavelength coverage, increasing total time required.}
\end{deluxetable}

\subsubsection{Supernovae and their hosts in the Wide, Fast, Deep (WFD) Survey}

While the smaller area of the deep fields ensures that host redshifts can be obtained for a higher fraction of all SN hosts than in the WFD, the wide field will yield a much larger number of SNe in total. 
This sample of hundreds of thousands of SNe will revolutionize cosmological analyses and enable extensive studies of systematics. The improvement is illustrated in the left panel of Figure~\ref{fig:sn_specsize}, which shows redshift histograms for the best-characterized SNe~Ia in the DDF and WFD, as compared to the current sample of known SNe. The LSST WFD survey will increase the number of SNe dramatically in the intermediate redshift range, complementing the deep sample from the DDF.  The 4MOST/TIDES program will target SNe and their hosts for spectroscopy across the WFD footprint, but will run only for the first half of the LSST survey \cite{tides}.


The distribution of SN host redshift measurements will depend on the allocation of spectroscopic resources from ground-based telescopes.  We have tested the impact on the Dark Energy Task Force Figure of Merit (FoM) from altering the fiducial redshift efficiencies from \cite{descsrd}, which are shown in Figure~\ref{fig:sn_specsize},
either by changing the total number of hours and simply rescaling the amplitude of the selection function, or by modifying the distribution to disfavor fainter hosts at higher $z$. 
From these tests we find that the DETF FoM is most sensitive to  the number of SNe at high redshift, and hence changes in the time allocated to the DDFs have the greatest effect.
High redshift completeness can be achieved across a broad range of galaxy populations for low-$z$ SNe in WFD, but that will likely be impossible for the higher-$z$ DDF sample.  This will make wide-area spectroscopy vital for investigating systematic uncertainties related to host galaxy properties, which may be a limiting factor in LSST SN cosmology.
%

\section{Transient Strong Lensing}

One of the most striking consequences of general relativity is that light from distant sources is deflected by the gravitational field of massive objects near the line of sight.
When the deflection produces multiple images, the phenomenon is known as ``strong gravitational lensing.'' 
The multiple images of strongly lensed sources arrive at different times because they  travel different paths and through different gravitational potentials to reach us \citep{refsdal64a,blandfordandnarayan92}.
When a strongly lensed source is time variable, the arrival time delays can be measured and combined with a mass model to yield cosmological constraints, particularly on the Hubble constant \h.  Strongly lensed QSOs, compound lens systems,  and strongly lensed supernovae each provide opportunities for measurement of cosmological parameters with LSST, and the LSST DESC plans to exploit them all \citep{descsrd}.

\subsection{Precision Cosmology with Strongly Lensed Supernova Time Delays}

\begin{figure}[htbp!]
	\centering
    \includegraphics[width=0.5\textwidth]{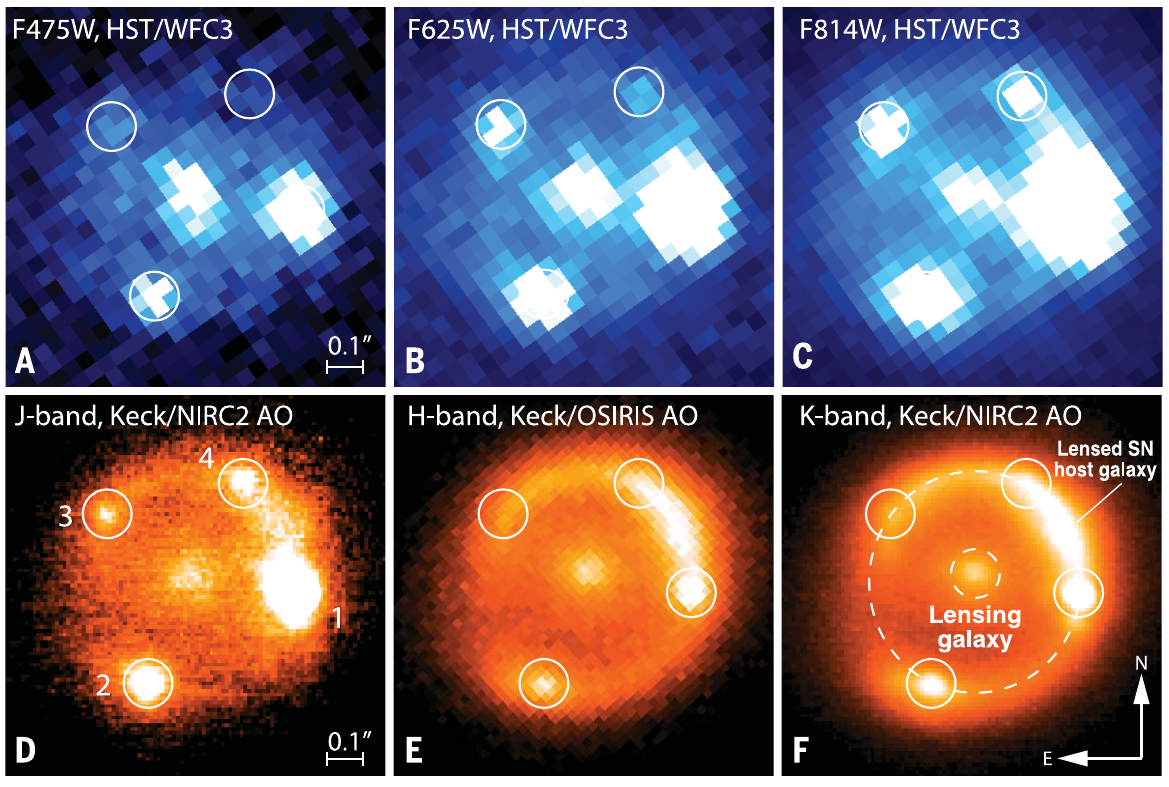}
    \caption{iPTF16geu, a Type Ia supernova at $z=0.4$ strongly lensed by an elliptical galaxy at $z=0.2$ (data from \textit{HST}\ and Keck). 
    This transient---the only known lensed \snia\ with resolved multiple images---was discovered by a predecessor of LSST, the intermediate Palomar Transient Factory (iPTF).
    The multiple images of the supernova are indicated with circles. 
    Due to the small image separation ($\sim$0.3$^{\prime\prime}$), high-resolution imaging was required to confirm that it was a strongly-lensed supernova.
    Figure reproduced from \cite{goobar16}.}
    \label{fig:geu}
\end{figure}

The prospect of cosmological constraints from strongly lensed supernovae was  illustrated with the ground-breaking Hubble Space Telescope observations of the first multiply-imaged lensed supernova, SN Refsdal \cite{kelly}. Ground-based time-domain optical imaging surveys similar in spirit to LSST, such as the intermediate Palomar Transient Factory (iPTF), played a key role in the discovery and follow-up observations of another strongly lensed supernova with resolved multiple images \citep[][see Figure \ref{fig:geu}]{goobar16}.

Time delays from lensed supernovae present opportunities to observe the earliest phases of supernova explosions, to infer cosmological parameters, and to map substructure in lens galaxies, but many more systems are needed to achieve these goals. Thanks to novel lensed-supernova hunting techniques, a new generation of alert-based wide-field imaging surveys that  began in mid--2018 with the Zwicky Transient Facility \citep[ZTF;][]{ztfnat} and will continue into the 2030s with LSST  will increase the size of this sample by orders of magnitude. 
ZTF is currently expected to yield $\sim$20 lensed supernovae over the course of its 3-year campaign, followed by thousands from LSST and WFIRST \citep{gnkc17}. 
This vast increase in sample size will enable groundbreaking new measurements with the potential to rapidly deliver precision constraints on the Hubble constant (\h) and dark energy.  The Hubble constant can be determined to exquisite precision with measurements of the \textit{nearby}\ universe  using the distance ladder \citep[][$\h=73.24\pm1.74\,\mpckms$]{riess16}. 
It can also be inferred with measurements of the \textit{primordial}\ universe using the cosmic microwave background (CMB), assuming a \lcdm\ cosmology \citep[][$\h=66.93\pm0.62\,\mpckms$]{planckhitens}. 
The tension between these local and distant measurements is palpable: they currently disagree by $\sim4\sigma.$ 
It is potentially a sign of new fundamental physics, such as sterile neutrinos or ``phantom'' dark energy \citep[e.g.,][]{phantom,freedman17,zhao17}, but could also be a sign of systematics in the measurements \citep[e.g.,][]{efstathiou14,trouble}.
Time delays between the multiple images of strongly lensed time-variable sources depend primarily on \h, and the mass distribution along the line of sight \citep{kochanek02, treu2010,fassnacht11,cosmography}.
Refsdal \cite{refsdal64b} first suggested using time delays from strongly lensed supernovae to  measure \h, but today, more than 50 years later, this measurement has yet to be made with precision.
Time delays from lensed supernovae have many advantages over time delays from strongly lensed quasars, which have been used to measure \h\ to 3.4\% in a \lcdm\ cosmology 
Lensed supernovae require 100 times less monitoring  (i.e., a few weeks rather than decades) and are less sensitive to microlensing \citep{gnkc17},  mass modeling systematics \citep{oguri03}, and selection bias \citep{collettcunnington16}.
Consequently, they provide the most direct and rapid route to sub-percent constraints on \h\ with strong lensing time delays.

\begin{figure}[htbp!]
	\centering
    \hspace{-37pt}
    \includegraphics[width=0.6\textwidth]{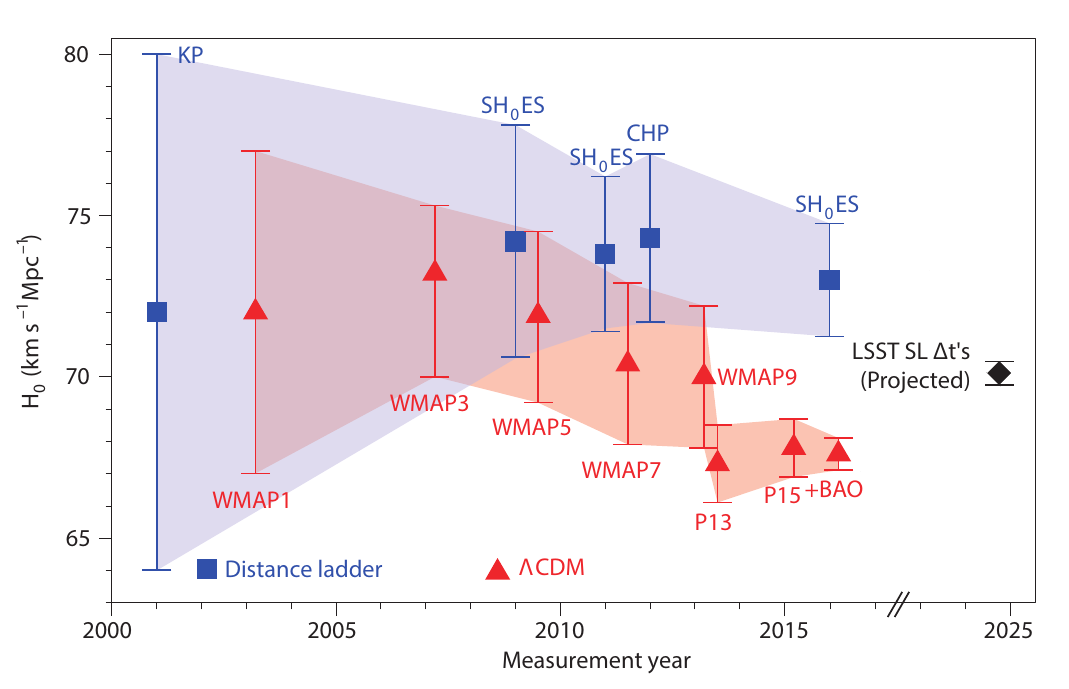}
    \caption{
    \footnotesize Examples of measurements of \h\ since the year 2000. Strong lensing time delays from systems discovered by LSST and observed with supplemental monitoring and follow-up spectroscopy as recommended in this white paper (black diamond) will resolve or increase the current tension in \h.
    Figure adapted from \cite{freedman17}.}
    \label{fig:impact}
\end{figure}

We forecast that \h~can be measured to sub-percent precision within \lcdm\ cosmological models with time delays from systems discovered in the first year of the LSST survey. The total number of supernova time delay systems over the LSST survey is expected to be $\sim 100$. Our projected constraint on \h\ as shown in Figure~\ref{fig:impact} is comparable in  precision to the leading current measurement from the combination of $\textit{Planck}$ and BOSS data \citep{aubourg15,planck15}, and it is almost ten times better than the current state-of-the-art constraints from quasar time delays \citep{Bonvin17}. In addition to constraining \h, time delays from lensed supernovae are sensitive to dark energy in a completely different way than cosmological probes based on distances and volumes \citep[e.g., the CMB, the \snia\  distance-redshift relation, BAO, and galaxy clusters;][]{linder04,linder11}, making time-delay measurements highly complementary. Adding in lensed supernovae from the first year of LSST increases the Dark Energy Task Force \cite{detf_report} figure of merit by a factor of 3 over an \snia-based constraint alone---a major gain.

{\bf Complementary methods:} The use of gravitational lens time delays as ``standard clocks'' to determine cosmological distances is complementary to the method of ``standard sirens'' using the gravitational wave signal emitted by merging neutron stars and black holes \cite{Snowmass2022:Tensions}.  That method requires redshift measurements for the host galaxies of the merger events. For binary neutron star mergers with kilonova couterparts its precision can be improved with spectroscopic observations of the optical-infrared (OIR) transient to constrain the merger orientation and reduce the orientation–distance degeneracy. Planned upgrades to gravitational wave detectors should greatly increase the samples for standard siren measurements, enabling 1\% measurement of $H_0$, but the more distant OIR transients will be exceedingly faint and may require 25--40~m-class telescopes for spectroscopic follow-up. Traditional ``standard candle'' distance measurements using the infrared tip of the red giant branch (IR-TRGB) method can also be extended to large samples of galaxies out to $\sim 100$~Mpc distances, well into the Hubble Flow, using deep, high angular resolution observations from 25--40~m-class telescopes, providing a third independent and complementary method for precision measurement of $H_0$ \cite{Astro2020:Beaton}.

\subsection{Photometric and Spectroscopic Follow-up of QSO Lens Systems}

LSST is expected to deliver hundreds of cosmologically useful lensed supernovae \citep{2018arXiv180910147G} and thousands of cosmologically useful lensed QSOs \citep{om10}. Initial observations of strong lenses piggybacking on wide-field multi-object spectroscopic surveys can help to characterize lensed QSO systems and identify the most useful ones.
However, measuring cosmological parameters to percent-level accuracy with strong-lens time delays requires two main ingredients once the ideal systems are identified. First, one must measure the time delays between the multiple images of a source, and second, the lensing potential must be inferred to convert the observed time delays into a time-delay distance. 

To measure time delays, high-cadence, high-resolution, multi-filter imaging of the resolved lensed images is required. In general, the LSST cadence may be insufficient for this purpose; as a result, dedicated follow-up imaging of lens systems with more frequent visits will be needed, but this can often be performed on smaller telescopes (2--4~m).
To model the lens systems, both  source and lens redshifts are required, as well as high-resolution imaging of the lens galaxy and lensed host galaxy to measure apparent positions with high precision. Kinematic velocity dispersions derived from lens galaxy spectra can also improve the models.  Adaptive optics (AO) IFU spectroscopy on 8--40~m-class telescopes (with aperture required depending on brightness) can satisfy all of these needs simultaneously, but the combination of AO imaging with slit spectroscopy would also suffice.  Additionally, multi-object spectroscopy of faint galaxies near the line of sight can help to resolve mass-sheet degeneracies and improve lens modeling. 

\subsection{Wide-field spectroscopy for lens systems}

LSST will discover $\sim 100,000$ strong gravitational lenses \citep{Collett15}, 100 times more than are currently known. This drastic change opens new opportunities to precisely constrain dark energy with lensed quasars \citep{Bonvin17}, lensed supernovae \citep{Goldstein18}, lenses with multiple background sources \citep{Collett14}, and lenses with a spectroscopic velocity dispersion \citep{Grillo08}, as well as by using strong lensing as a calibration tool for weak lensing shear \citep{Birrer18}. Each of these science cases requires redshifts for lens and source, both to confirm candidates as lenses and to convert lensing-derived quantities into cosmologically meaningful constraints. The LSST lenses will be uniformly distributed across the LSST footprint, with typical $r$-band magnitudes of $20.1 \pm 2.4$ for the lens and $23.7 \pm 0.7$ for the source. As a result, many lenses are bright enough for redshift measurements via targeted fibers within very wide-area surveys, enabling identification of the systems best suited for follow-up as described above.  



\section{Peculiar Velocities Measured From Supernovae}
In the upcoming decade cadenced wide-field imaging surveys 
will increase the number of identified   Type~Ia supernovae (SNe~Ia) at $z<0.2$  from  hundreds up to one hundred
thousand.  The increase in the number density  and solid-angle coverage 
of SNe~Ia, in parallel with improvements in the standardization of
their absolute magnitudes, now make them competitive probes of the growth of structure and hence of gravity.  Each SN Ia can have a peculiar velocity $S/N \sim 1$ with the distance probative power of $30-40$ galaxies, so a sample of 100,000 SNe Ia is equivalent to a survey
of 3 million galaxies, in a redshift range where we currently have 
$\lesssim$100,000 galaxies. The peculiar velocity power spectrum
is sensitive to
the effect of gravity on the linear growth of structure.
Measurements with 4--5$\sigma$ confidence will be possible in the coming decade by exploiting the large numbers of $z < 0.09$ SNe Ia now and soon to be discovered \citep{2014ApJ...788...48S, 2018PASP..130f4505T, 2020arXiv200109095G}  together with even more out to $z=0.2$ and beyond to come from the Vera C. Rubin Observatory LSST~\citep{2018arXiv180901669T}.
These constraints 
have the same statistical significance as
those from DESI or 
Euclid, but are measured essentially at a given cosmic time 
rather than averaged over a broad range $z\sim$1--2, and at 
low redshift where modifications of gravity may be most apparent. 
Together, SNe~Ia and high-redshift peculiar velocity tracers are sensitive to physics beyond the standard $\Lambda$CDM cosmology.

Given the anticipated stream of fresh SNe, a coordinated network of follow-up facilities is
then required to 
determine their peculiar velocities.
The signals the network must provide are deviations from the average Hubble expansion:

{\bf Early-Phase Screening:}
Cadenced wide-field imaging surveys discover SNe~Ia together with a background of other transient events.
Though not strictly required, screening enables the efficient use of limited telescope resources by identifying
a  subset of likely active SNe~Ia targets for follow-up at peak brightness.
Screening is primarily based on the temporal and color evolution during the early post-discovery phases of transient light curves.

{\bf SN~Ia Classification:}    
Pure SN~Ia samples have
lower absolute magnitude dispersion, and hence better velocity precision, relative to samples that suffer from non-Ia contamination.
SNe~Ia are defined based on their spectroscopic features,
so spectroscopic coverage of the relevant features ensures the purity of the analysis sample.  
Photometric classification dilutes the signal in the scatter of the Hubble diagram, while selection from
only early-type hosts has core-collapse contamination and significantly reduced sample size.

{\bf Host-Galaxy Precision Redshifts:} Peculiar velocities come from the difference between host-galaxy and cosmological redshifts.  Fractional
host-galaxy redshift uncertainties from moderate-resolution spectroscopy ($R>100$) 
do not contribute significantly
to the error budget.   In many cases the redshifts could come naturally
from the galaxy signal in  the spectroscopic transient classification.  Alternatively, 
moderate-resolution bright-galaxy  redshift surveys  can  provide multiplexed observations of 
a significant fraction of nearby SN hosts before discovery or after transient light has faded away.

{\bf Precise SN~Ia Absolute Magnitudes:} SNe~Ia are standardizable candles, in that their absolute magnitudes can be inferred from intrinsic observables such
as multi-band light-curve shapes, spectroscopic features, and host-galaxy properties. 
For the same supernova, the absolute magnitude uncertainty derived from a dataset with sparse light curves, a small number of filters, limited wavelength coverage, and
lacking spectral-feature measurements is larger than the dispersion derived from a broader dataset with dense light curves, a large number of filters, 
broad wavelength coverage, and spectral-feature measurements.  Surveys with three well-sampled rest-frame optical light curves obtain $\sim 7$\%\citep{2018ApJ...859..101S} distance uncertainties; 
standardization studies show that supplemental
infrared data \cite{2012MNRAS.425.1007B,2018ApJ...869...56B} or spectrophotometry at peak brightness
\cite{2015ApJ...815...58F,Boone2020} can  yield distances as good as $\lesssim 4$\%. 
Follow-up measurements
that give this better precision yield SNe with  3$\times$ the peculiar-velocity probative power compared to those with only optical light-curves.

{\bf SN~Ia Observed Magnitudes:}
The difference between observed
and absolute magnitudes gives the distance to a supernova.  Generally the
observed magnitudes come from the amplitudes of the
same multi-band light curves whose shapes and colors are used to infer absolute magnitudes.

Two reference surveys are introduced to assess the supplemental resources needed for a complete SN peculiar velocity program.  The first ``pathfinder''
survey corresponds to a  sample of $\sim 5000$, $g<19$~mag, $z<0.09$ SN~Ia
discoveries as would be discovered by
ASAS-SN+ATLAS, ZTF-II during a three-year survey.
The second ``legacy'' survey is composed of a sample of $\sim$150,000, $r<20.5$~mag, $z<0.2$ SN Ia as will be discovered by LSST over the course of ten years.

Nominal follow-up networks that collect the requisite peculiar-velocity data
for the above
reference surveys are now described.
The follow-up strategies presented
have
the advantage of being based on successful precedents, but do not represent the only viable strategy.

{\bf Wide-field ($\gtrsim 1$ sq.\ deg.) Imaging Survey on a $\lesssim 2$m Telescope:}
Early-phase screening ideally would come from this search.  The observing strategy of ZTF (and presumably ZTF-II) is highly cadenced and so provides good early photometric classification\citep{2019PASP..131k8002M}. However, the original LSST Wide Fast Deep (WFD) survey produces sparse per-band light curves with little early-time data and poor distance determinations \citep{2018arXiv181200515L}.
    The ultimate LSST WFD
    strategy is currently unspecified; hence its capability for early typing remains unknown. The Rubin Observatory has convened the Survey Cadence Optimization Committee, which is 
    currently slated to decide on the observing strategy by late 2022\footnote{See \url{https://www.lsst.org/content/charge-survey-cadence-optimization-committee-scoc}}.
    This source of risk is mitigated by planning for a wide-field imaging survey on a different telescope designed to 1) fill in the temporal gaps of the WFD survey to enable early classification; 2) implement an observing strategy optimized for low redshift peculiar velocities, e.g.\ favor larger solid angle and shallower depth than envisioned for WFD.
    The LS4 project\citep{LOI-LS4}, a next-generation transient search on the  La Silla Schmidt Telescope with an upgraded camera, satisfies these needs.  Other options include the  VST/OmegaCam\cite{2015A&A...584A..62C} and DECam.
    
{\bf Targeted (IFU) Spectroscopy of Active SNe:}  Spectroscopic follow-up with moderate spectral resolution of
    likely SNe~Ia provides 1) classification; 2) host-galaxy redshifts; 3) precision SN~Ia absolute magnitudes.
    The SNIFS Integral Field Unit (IFU) spectrograph \citep{2004SPIE.5249..146L} mounted on the UH-88" is a proof of concept, having generated a time series of $\sim 300$ SNe~Ia\citep{Aldering2020} at $z<0.08$. 
    (Other IFU designs\citep{2010SPIE.7735E..08B, 2019BAAS...51g.198B} are also viable.) The IFU allows for the subtraction of a host-galaxy reference  yielding a spectrophotometric SN-only spectrum.
    For the ``pathfinder survey'', a 2.2~m telescope requires an average 20 minute exposure 
    to measure the spectral features of one SN~Ia near maximum light, which translates to $\sim 160$ clear nights per year for three years to get the full sample.  Total clock time includes overheads for poor weather and non-Ia contamination.
    In addition, a comparable amount of non-time-critical reference spectra must be obtained at least a year after the SN light has faded. 
    The ``legacy'' survey of 150,000 SNe Ia requires  added 4m-class telescopes with 7$\times$ the open-shutter time per year (3 telescope years per year plus weather overhead) for ten years plus additional time for references.
    
{\bf Targeted Spectroscopy of Host Galaxies:} The redshifts of a subset of hosts will not be accessible from the IFU data. In these cases redshifts of host galaxies 
can be efficiently obtained using 
wide-field multi-object spectroscopy with DESI\citep{2016arXiv161100036D} or 4MOST\citep{2019eeu..confE..56S} even
after the supernova light
has faded.

There are other approaches for designing  the network that can be evaluated during the Snowmass Process.
One option is to  focus on NIR photometry to improve
SN~Ia distances, e.g.\   at UKIRT.
Another is to decouple the classification and redshifting instruments by obtaining (sub)classifications using $R<100$ 
spectroscopy that cannot deliver a precision redshift, e.g., SED Machine\citep{2018PASP..130c5003B},  
and getting redshifts from dedicated
host-galaxy spectroscopy;
indeed this approach is being applied to the shallow discoveries of ZTF\citep{2020ApJ...895...32F}.
Another option is to use
photometry only, without supplemental spectroscopy, to calibrate SN absolute magnitudes; the probative power of each SN drops significantly in this case, but still remains much better than competing Fundamental Plane and Tully-Fisher distances.  Similarly, follow-up of a reduced subset of SNe with scaled-back follow-up resources, while not taking full advantage of all the discoveries, can
still outperform galaxy-derived distance surveys.
\section{Photometric redshift training and calibration}

Photometric redshifts -- i.e., redshift estimates obtained from imaging data alone -- will be critical for all LSST probes of cosmology.  The cosmological tests enabled by LSST all rely on determining the behavior of some quantity with redshift, $z$, but we cannot measure spectroscopic redshifts (spec-$z$'s) for the large numbers of objects detected in the LSST imaging data.    Even in cases where follow-up spectroscopy of individual objects will be needed (e.g., strong lens systems and some supernova studies), photo-$z$'s are used to identify targets of interest.  However, if photometric redshift estimates are systematically biased, dark energy inference can be catastrophically biased as well (see, e.g., \cite{hearin_etal10}); as a result photo-$z$'s are both a critical tool and a major source of concern affecting all cosmological analyses.  The great depth of LSST data is a compounding factor, as it makes the challenges associated with obtaining follow-up spectroscopy to the depths of the samples used for LSST more difficult.

Lacking a comprehensive knowledge of galaxy evolution, the only way in which photo-$z$ errors can be reduced and biases characterized is via robust spectroscopic redshift measurements of a set of galaxies.  We follow \cite{specneeds} in dividing the uses of spec-$z$'s into two  classes, ``training'' and ``calibration.'' 
 {\bf Training} is the use of samples with known $z$ to develop or refine algorithms, and hence to \textit{reduce  random errors on individual objects' photo-$z$'s}: i.e., to improve the \textit{performance} of algorithms at estimating the redshift of a galaxy.  Photometric redshifts that are trained from larger and more complete spectroscopic samples and hence have smaller uncertainties could greatly improve the constraining power of LSST, for example by providing sharper maps of the large-scale structure 
 and improved clustering statistics, providing better photometric classifications for supernovae, enabling identification of lower-mass galaxy clusters at higher confidence, and yielding better intrinsic alignment mitigation for weak lensing measurements.  If photo-$z$ accuracies are limited only by photometric errors (as would be true if a perfect -- i.e., very large and totally unbiased -- training set could be obtained), it has been found in simulations that LSST can deliver photo-$z$ estimates with sub-$2\%$ uncertainties ($\sigma_z < 0.02(1+z)$).  However, errors in redshifts from real data sets at LSST depth with our current knowledge of galaxy spectral energy distributions are closer to 5\%.  Achieving the ideal performance by having a large training sample spanning the properties of objects used for cosmology would improve the Dark Energy Task Force Figure of Merit from LSST lensing and Baryon Acoustic Oscillations alone by $\sim 40\%$ \cite{detf,2006JCAP...08..008Z}.

{\bf Calibration} is the problem of determining the true overall $z$ distribution of a sample of objects; i.e., the \textit{characterization} of redshift distributions that will be used in analyses.  Miscalibration will lead to {\it systematic} errors in photo-$z$'s and hence downstream analyses \cite{2006ApJ...644..663Z,2006JCAP...08..008Z,2006astro.ph..5536K,tysonconf,hearin_etal10}.  
Photo-$z$ calibration requirements for LSST are extremely stringent \cite{descsrd}.  If photo-$z$'s are systematically biased in an unknown way or if their errors are poorly understood, dark energy inference will be biased as well (e.g., because we are effectively measuring the distance to a different redshift than is assumed in calculations).  
%

For both training and calibration, we need sets of galaxies for which the true $z$ is securely known; this generally requires medium or high-resolution spectroscopy (low-resolution spectroscopy or many-band imaging yield less-secure redshifts with high incorrect-redshift rates, which can have a catastrophic impact on the training of algorithms and the characterization of redshift distributions).  If spec-$z$'s could be obtained for a large, unbiased sample of objects down to the faint limit of the samples to be used for LSST cosmology, both needs could be fulfilled using the same data; we describe a sample which could optimistically serve this dual purpose in \autoref{photoz_deep}. However, many faint galaxies in existing spectroscopic samples fail to yield secure redshift measurements.  Photometric redshift performance will still improve if only incomplete spectroscopic samples are obtained, but the accuracy of the characterization of redshift distributions would be compromised. In that case, other methods may be needed for the calibration of redshift distributions.  Complementary spectroscopic datasets that can serve this purpose are described in \autoref{photoz_wide}.  

\subsection{Deep spectroscopy for improving photometric redshift performance}

\label{photoz_deep}

\begin{figure}[t]
\centering
\includegraphics[width=0.85\textwidth]{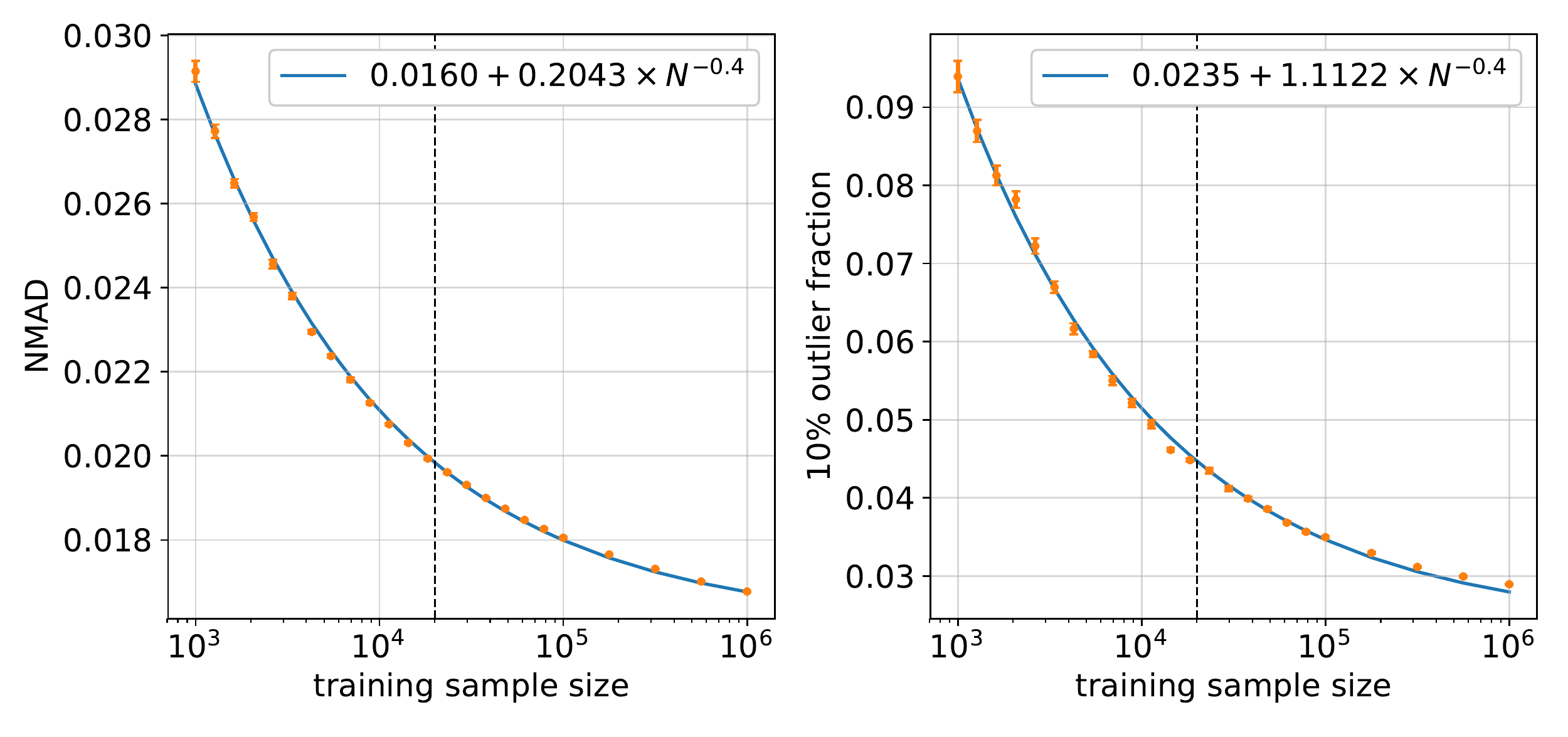}
\caption{Orange points show photometric redshift errors and outlier rates versus the number of galaxies in the training set for galaxies with simulated LSST photometric errors.   Photo-$z$'s were calculated using a random forest regression algorithm. The left panel shows the photo-$z$ error, quantified by the normalized median absolute deviation (NMAD) in $(z_\text{phot}-z_\text{spec})/(1+z_\text{spec})$, as a function of training set size; similarly, the right panel shows the fraction of 10\% outliers, i.e. objects with $|z_\text{phot}-z_\text{spec}|/(1+z_\text{spec})>0.1$. A vertical dashed line shows the sample size for the baseline training survey from \cite{specneeds}.  The blue curves represent simple fits to the measurements as a function of the training set size, $N$. This analysis uses a set of simulated galaxies from Ref.~\cite{Graham2018} that spans the redshift range of $0<z<4$, using a randomly-selected testing set of $10^5$ galaxies for estimating errors and outlier rates; these catalogs are based upon simulations from Refs. \cite{grahamcat1},\cite{grahamcat2}, and \cite{grahamcat3}.}
\label{fig:training_size}
\end{figure}

In a white paper developed for the previous Snowmass process \cite{specneeds}, it was concluded that an effective training set of photometric redshifts for the LSST weak lensing sample would require highly-multiplexed medium-resolution ($R \sim 4000$) spectroscopy covering as much of the optical/infrared window as possible with very long exposure times on large telescopes. To enable photo-$z$ direct calibration errors to be subdominant to other uncertainties if the training set were used for that purpose, the spec-$z$ sample must comprise at least 20,000 galaxies spanning the full color and magnitude range used for cosmological studies, reaching $i=25.3$ for LSST.  As can be seen in Fig.~\ref{fig:training_size}, improvements in photo-$z$ errors and outlier rates are slow beyond this point, so larger samples have limited utility for improving the performance of algorithms as well as for calibration.

A critical issue for machine learning-based photometric redshift  algorithms 
\cite[e.g.,][]{2015MNRAS.452.3100C} is that sample/cosmic variance in the regions with spectroscopy can imprint on the inferred redshift distribution of objects over the whole sky, biasing photo-$z$ results.  
To both quantify and mitigate this effect, the survey strategy described in Ref.~\cite{specneeds} seeks to obtain spectroscopy spanning at least 15 widely-separated fields a minimum of 20$^\prime$ in diameter.  
Such a survey has comparable sample/cosmic variance to the Euclid C3R2  strategy of six 1 sq.~deg.~fields \cite{2017ApJ...841..111M}, but requires only $\sim$ 22\% as much sky area to be covered; by spanning more fields than C3R2 it also allows more robust identification of regions that are overdense or underdense at a given $z$.  

We have calculated the amount of dark time required for a survey for training LSST photometric redshifts for a variety of instruments and telescopes in Table \ref{table:photoz_times}, updated from the results presented in Ref. \cite{specneeds}.  In this table we calculate the total time required to obtain spectroscopy of objects down to magnitude $i=25.3$ with the same signal-to-noise that would be achieved with the Keck/DEIMOS instrument for objects with $i=22.5$, for a total sample of at least 20,000 objects distributed across 15 fields of at least 20 arcminutes in diameter, 6 fields of at least 1 degree in diameter, or 4 fields of at least 2 degrees in diameter (in order to mitigate the effects of sample/cosmic variance).  Only instruments or combinations of instruments with the capability of obtaining spectra over the full optical window with sufficient spectral resolution to resolve the [OII] 3727 angstrom doublet are considered; with lower resolution, it is infeasible to obtain highly-secure redshifts for most galaxies at $z>1$.   Results are given in units of ``dark-years'': one dark-year consists of 365 8-hour-long dark nights, of which one-third are lost to weather and instrumental overheads.



We note that such a survey has strong synergies with studies of galaxy evolution, offering the possibility of co-funding that extends beyond HEP.  It would determine the range of galaxy SEDs (and hence star formation histories) as a function of their local environment across the redshift and magnitude range covered by LSST cosmology samples (reaching down to $\sim L_*$ at $z=2$, and including brighter galaxies to $z=3$).  Improved photo-$z$'s will also enhance a variety of galaxy evolution science from LSST, making the results useful to broader communities.

\begin{deluxetable}{lllll}
\tabletypesize{\footnotesize}
\tablecaption{Time required for photometric redshift training spectroscopy}
\label{table:photoz_times}
\tablehead{ \colhead{Instrument / Telescope} & \colhead{Collecting}  & \colhead{Field area}  & \colhead{Multiplex} & \colhead{Total time} \\
 & \colhead{Area (sq. m)} & \colhead{(sq. arcmin)} &  & \colhead{(dark-years)}\\ \hline}
 \startdata
  4MOST & 10.7 & 14,400 & 1,400 & 1.4 \\ 
        Mayall 4m / DESI & 11.4 & 25,500 & 5,000 & 1.4 \\ 
        WHT / WEAVE & 13.0 & 11,300 & 1,000 & 1.6 \\ 
        Megamapper (Magellan-like) & 28.0 & 25,416 & 20,000 & 0.6 \\ 
        Subaru / PFS & 53.0 & 4,500 & 2,400 & 0.4 \\ 
        VLT / MOONS & 58.2 & 500 & 500 & 2.7 \\ 
        Keck / DEIMOS & 76.0 & 54 & 150 & 6.8 \\ 
        Keck / FOBOS & 76.0 & 314 & 1,800 & 0.8 \\ 
        ESO SpecTel & 87.9 & 17,676 & 15,000 & 0.2 \\ 
        MSE & 97.6 & 6,359 & 3,249 & 0.2 \\ 
        GMT/MANIFEST + GMACS & 368.0 & 314 & 420 & 0.5 \\ 
        TMT / WFOS & 655.0 & 25 & 100 & 1.2 \\ 
        E-ELT / Mosaic Optical & 978.0 & 39 & 200 & 0.5\tablenotemark{1} \\ 
        E-ELT / MOSAIC NIR & 978.0 & 46 & 100 & 0.8 \\ 
\enddata
\tablenotetext{1}{For E-ELT, observations in both the optical and near-IR settings are required to achieve the required wavelength coverage, increasing total time required.}
\end{deluxetable}

\subsection{Deep spectroscopy for testing the impact of blending on photometric redshifts}

\label{photoz_blends}


Due to its unprecedented depth and sensitivity to the low-surface-brightness outer regions of galaxies, the probability of two or more sources overlapping in LSST data is high. 
Approximately 63\% of LSST sources will have at least 2\% of their flux coming from other objects, in contrast to $\sim$30\% of  sources in Dark Energy Survey (DES) data~\citep{Samuroff18,Sanchez19}. 
These overlaps complicate the measurement of galaxy fluxes and shapes, requiring accurate deblending \citep{Melchior18,Euclid19} or statistical correction techniques. Any residual light contamination will result in biases to individual-object photo-$z$'s or potentially even in estimates of overall redshift distributions~\citep{Gruen18}.

Deep multi-object, medium-resolution spectroscopy can detect the presence of blends and the redshifts of each component by identifying superimposed features in a spectrum, providing new training samples for deblenders and constraints on statistical corrections for deblending effects.  While many blends can be detected as having multiple components in space-based data, some are so close that only spectroscopy can provide a definitive indication of a problem.  As a result, the proposed photometric redshift training spectroscopy should also greatly enhance studies of deblending down to the depth of the LSST Gold sample, $i_\text{lim}=25.3$~\citep{descsrd}.



\subsection{Wide-area spectroscopy for characterizing photometric redshift distributions}

\label{photoz_wide}

Given the level of accuracy with which the redshift distributions of LSST photometric samples must be known, if we wish to characterize them with a representative sample of galaxies such as would be obtained from the spectroscopy described in \autoref{photoz_deep}, that spectroscopy would need to have an extremely high ($>99\%$) spectroscopic redshift success rate (i.e., the fraction of spectroscopic targets for which a secure redshift is obtained) and an extremely low ($<1\%$) incorrect-redshift rates (i.e., the fraction of putatively secure redshift measurements which are not in fact correct).  Existing deep spectroscopic samples have fallen well short of these goals, meeting the incorrect-redshift requirement only for the most robust (and hence most incomplete) subsets of redshift measurements \cite{Newman2013,Lilly2007}.  Based on the systematic incompleteness of existing deep surveys, we therefore proceed under the assumption \citep[e.g., as applied in][]{specneeds} that direct calibration via a large, representative spec-$z$ sample will not be possible at the depth of LSST weak lensing samples.

However, methods based upon cross-correlating the locations on the sky of LSST galaxies with the positions of spec-$z$ samples as a function of the spectroscopic $z$ 
(sometimes referred to as ``clustering redshifts'') 
can provide an accurate photo-$z$ calibration for dark energy applications  \cite{2008ApJ...684...88N}. These techniques exploit the fact that all galaxies cluster together on large scales, so those objects with spec-$z$'s intrinsically cluster on the sky only with the subset of galaxies in other samples that are at nearly the same redshift, but not with closer or more distant objects.
We can use this to determine the true redshift distribution of objects in an unknown sample \citep[e.g.,][]{Schmidt:2013sba,Rahman:2014lfa} as input to dark energy measurements.  Precise calibration only requires that the spectroscopic samples span the redshift range of the LSST objects and cover a wide area of sky.  

A spectroscopic sample containing of order $\sim$$10^5$ objects spanning thousands of square degrees and the full redshift range of LSST photometric samples and covering a fair sample of observing conditions, extinction, etc. should be sufficient for calibrating LSST photometric redshifts \cite{specneeds}.  
The DESI survey \cite{2016arXiv161100036D} which began in 2021 should meet the sample size and redshift range requirements \cite{specneeds} for this application: it will overlap with at least 4000~deg$^2$ of the LSST footprint, obtaining more than 10 million redshifts of galaxies from $0<z<1.6$ and quasars at $z < 4$ over that area.  
However, because DESI is located at the northern end of the LSST 
footprint, there is a risk of the LSST photometry -- and hence the photo-$z$'s to be calibrated -- being systematically different in this region (e.g., due to difficulties in correcting for the atypical distribution of airmasses). 
As a result, greater southern coverage is extremely desirable.  The 4MOST survey, which will use a dedicated wide-field multi-object spectrograph on the VISTA telescope at Paranal Observatory \cite{4most}, plans to obtain samples with DESI-like target selection (but lower number density) over $\sim$1000 deg$^2$ for Emission Line Galaxies (ELGs) and $\sim$7500 deg$^2$ for Luminous Red Galaxies (LRGs) and quasars; these areas may overlap with DESI.  It would thus be valuable to enlarge the samples available for cross-correlation calibration in the southern part of the LSST footprint; this would require new survey data in regions of sky not covered by the DESI survey.  A follow-on survey with the DESI instrument that covers southern regions targeted by LSST but not the DESI Legacy Survey imaging could potentially serve this purpose, or alternatively samples from a high-multiplex instrument located further to the south.



\section{Intrinsic Alignments}

Intrinsic correlations between shapes of galaxies that are physically near each other (``intrinsic alignments’’  or IA) are a known contaminant to precision weak gravitational lensing (WL) measurements -- see, e.g., the reviews \cite{joachimi15rev,Troxel:2014dba} and references therein. If not accurately accounted for, their presence can generate significant biases in cosmological analyses \citep{Kirk12, Krause16, blazek19,Yao17}. More recently, it has also been recognized that through selection effects, IA can contaminate redshift-space distortion (RSD) analyses in DESI and other future spectroscopic galaxy clustering projects \cite{martens18}.

There are two types of IA correlations. The first happen when considering pairs of galaxies that are both within the same large-scale structure, thus directly inducing intrinsic shape correlations known as ``II'' type. The second type is due to the fact that a given matter structure will both align physically nearby galaxies and shear background galaxies. The resulting correlations are known as ``GI'' type. The latter is particularly problematic as it is present even between galaxy bins significantly separated along the line of sight and thus cannot be easily removed, even with precise redshift information. It is this latter effect that can also impact RSD studies, since the apparent impact on galaxies will depend on the angle with the observer's line of sight.

Mitigation of IA typically requires jointly modeling the effect along with the cosmological lensing signal as well as other astrophysical and observational effects.\footnote{``IA self-calibration’’ \citep[e.g.][]{Yao17} is an alternative approach that removes IA without explicitly marginalizing over model parameters. However, constructing the relevant data vectors still requires an accurate understanding of IA correlations.}
Improved understanding of the underlying IA behavior will allow us to robustly perform this modeling without significantly degrading the overall constraining power of the analysis. Indeed, to optimize final statistical uncertainties, IA modeling should have as few parameters as possible, and these parameters will ideally have strong priors from targeted observational studies. Moreover, as our understanding of IA improves, we will not only be able to effectively remove them as a contaminant, but also extract additional cosmological information from the shape correlations (e.g.\ \cite{Chisari16}).

Despite this clear need for observational IA constraints, to date it has been challenging to obtain measurements with sufficient signal to noise, especially for the samples most relevant to future lensing projects such as LSST. Because IA and weak lensing can mimic each other, and both are weak signals compared to the intrinsic ``shape noise'' of galaxies, IA measurements are most effective when precise 3D locations of galaxies are known, i.e.\ when spectroscopic-quality redshifts have been obtained.
Below, we briefly describe how both deep/narrow and wide/shallow MOS data in the future can enable significantly improved constraints on Dark Energy, and cosmology more broadly, in the presence of IA.


\subsection{Deep spectroscopy for direct tests of intrinsic alignments}


By enabling the 3D localization of galaxies, a deep MOS campaign would provide greatly-improved direct constraints on intrinsic alignments for {\it typical} weak lensing sources, rather than only for the bright and nearby objects which current datasets constrain \citep{Joachimi11,Samuroff18,Johnston18}.  
Such data would extend our knowledge of IA to unexplored regimes in galaxy parameter space, resolve the current inconsistencies between predictions of different hydrodynamical simulations \citep{Tenneti15,Chisari15,Chisari16}, and allow better priors to be placed on IA model parameters. All of these advances will improve the cosmological constraining power of future lensing studies, including LSST \cite{Krause16,Yao17,fortuna21a}.


Currently, the faintest magnitude-limited sample with a significant IA detection has $i_\text{lim}<19.8$ \cite{Johnston19}. Forecasts based on recent measurements \cite{Singh14,Johnston19} indicate that meaningful IA constraints require $\gtrsim 10^5$ galaxies with measured shapes and spectra. Obtaining this many spectra for a representative sample at the magnitude limit of the LSST Gold sample ($i_\text{lim}=25.3$) would require a significant expansion of the photo-$z$ training program described above, and may be infeasible with currently planned facilities.
However, even if we cannot reach magnitudes as faint as $i_{\rm lim}=25.3$, extending direct IA tests closer to the LSST limits would be very valuable. 
Most of the IA signal will come from scales where shape noise dominates, and thus total IA constraining power (for a fixed number of galaxies) is maximized if the surface density of targets is high. Such a dense sample can be obtained by switching out targets to other, brighter galaxies during the photo-$z$ training survey once secure redshifts are obtained; in this way, $\gtrsim 10^5$ spec-$z$'s for  objects with $i \lesssim 24$ should be obtainable during the training survey (only 9\% as much observing time is needed to obtain the same S/N at $i=24$ as at $i=25.3$). Additional information will come from cross-correlations with shallower, wider-area surveys, as described below.


\subsection{Wide-area spectroscopy for cross-correlation studies of intrinsic alignments}
 %
%
%
Wide-field MOS will enable measuring the cross-correlation between positions of bright galaxies with spectroscopy and intrinsic shapes of fainter galaxies used for lensing analyses, constraining IA models with greater precision than possible with photometric data alone. Such measurements can place upper limits on IA or constrain parameters associated with IA models \citep[e.g.,][]{2014MNRAS.445..726C,Singh14,Johnston19}.
Wide MOS over a contiguous area would also enable cosmic web reconstruction \citep[e.g., as in][]{Kraljic18}, which could be helpful in testing some of the predictions of cosmological simulations \citep{Codis18}. 
Furthermore, cross-correlations using the spec-$z$ sample will allow us to break degeneracies between IA parameters and photo-$z$ uncertainties. 
%
%
%
Current measurements from spectroscopic samples \citep{Singh14,Johnston19} have shown that wide-field samples of $\sim$ 100k--500k objects with both shapes and spec-$z$ measurements can place useful constraints on IA amplitude and scale dependence. 
The LSST-DESI overlap should enable measurements which greatly surpass what is currently possible, especially if northern coverage by LSST is increased (as proposed in \citep{BigSky}). 4MOST galaxy samples will further enhance these analyses. However, larger samples over wider areas would yield better constraints on IA models, improving S/N in IA measurements by up to $\sim2\times$ if samples could be expanded to cover the full LSST extragalactic footprint. This would likely require a new high-multiplex facility in the Southern hemisphere.  Smaller-field surveys of fainter galaxies, described above, will provide complementary constraints on IA. 

Intrinsic alignments have also been proposed as a probe of cosmology and novel physics in their own right \citep{Schmidt15,Chisari16}, and a wide MOS campaign would improve their performance for exploring primordial non-Gaussianity, potential new interactions in the dark sector, and as a multi-tracer probe of large-scale structure.

\section{Galaxy cluster studies}

The observed abundance of galaxy clusters as a function of their mass provides a sensitive probe of the growth of structure and, to a lesser degree, the expansion history of the universe.  However, galaxy cluster abundance measurements from LSST will require additional data to mitigate systematic effects; as an example, the accurate calibration of the relationship between observable cluster properties and their total mass must be very well-understood.  

\subsection{Deep spectroscopy in galaxy cluster fields}

Deep multi-object spectroscopy of galaxies in a set of fields containing galaxy clusters will improve LSST cluster cosmology in a number of ways, while simultaneously resolving open questions about galaxy evolution by determining differences between SEDs of galaxies in clusters versus the field, potentially leading to interest from outside HEP in joint projects. 

{\bf Training and Testing Photometric Redshifts in Cluster Fields:} Photometric redshifts are a critical tool for calibrating the masses of galaxy clusters via weak gravitational lensing. Photo-$z$'s are already a leading source of systematic uncertainty in current work \citep{WtG3,McClintock19}, but the requirements on photo-$z$ accuracy for LSST will be even more stringent\citep{descsrd}. Photo-$z$'s are vital for distinguishing the lensed background galaxy population from the unlensed foreground and cluster population; unaccounted-for contamination of the background will lead to underestimation of cluster masses.  
However, photo-$z$ performance may degrade in higher-density regions due to the differing galaxy populations of clusters vs. the field, magnification and reddening of background sources, and severe blending due to cluster galaxies \citep[e.g.][]{CLASH_photoz}.  

Photo-$z$ algorithms are generally trained and evaluated on fields selected not to contain massive structures.  
However, there is reason to expect that photo-z algorithms may not perform as well in areas of the sky that host clusters as in the field. The incidence of blending is worse in such areas due to the denser environment and contamination by intra-cluster light, causing larger errors in the photometry.  
Lensing magnification changes the observed brightness and size of background galaxies while the mix of early- vs. late-type galaxies is different in clusters than in the field; these effects may lead to mismatches with photo-$z$ priors.  
Furthermore, due to the profound impact of the cluster environment on galaxy evolution, clusters might host galaxy types that are very rare in the field. They also may host dust which would cause reddening of background galaxies.  

As a result, it will be important to quantify the performance of photo-z algorithms in cluster fields specifically.  Doing so requires comprehensive multi-object 
spectroscopy down to weak lensing depths ($i \sim 25.3$) for a sample of $\sim 20$ clusters spanning a range of redshifts. This is best achieved with high-throughput, high-multiplex spectrographs with FOVs of $\sim 10^\prime$ (wide-format IFUs may be suitable in cluster cores) on the largest telescopes available; the multi-object spectrographs planned for the US-ELT facilities would be well-suited for this work.  

Such a program would be able to characterize the performance of photo-$z$ probability distributions as a function of magnitude, redshift, and cluster properties.  A sample of 1000 objects per cluster would enable a crude binning into 3 magnitude and 3 redshift bins with $
\sim$100 galaxies per bin, enough for a statistically meaningful evaluation of photo-$z$ performance at the percent level.  To cover a range of cluster redshifts, masses, and dynamical states, $>20$ clusters should be surveyed.  It is critical to achieve near-complete redshift success rates to avoid biases from target populations for which no redshift is measured in a first attempt.  Such data will also yield very valuable insight into LSST deblending performance in general.

{\bf Measurements of Cluster Kinematics and Infall Velocities:}
Deep multi-object spectroscopy in fields hosting clusters will also enable direct mass estimates for galaxy clusters via the infall method \citep{Diaferio99,Rines03,Falco13,Arthur17},  providing an additional calibration of the mass-richness relation \citep{Rines18} ``for free" from the photometric redshift training/test spectroscopy.  We hope to obtain $\sim 200$ redshifts at projected separation $< 5$ Mpc for each cluster targeted for photo-$z$ studies, providing purely kinematic mass estimates.  In conjunction with weak-lensing measurements, spectroscopy of halo infall regions also enables sensitive comparison between the dynamical and the lensing potential of gravity, which can differ from each other at measurable levels in several modified gravity theories that seek to provide an alternate explanation for cosmic  acceleration \citep[e.g.,][]{Zu14}.  

%
%


\subsection{Additional constraints from Sunyaev-Zel'dovich measurements}

Planned CMB surveys such as CMB-S4~\cite{CMB-S4:2016ple,Abazajian:2019eic}, will detect a very large number of galaxy clusters through the thermal Sunyaev-Zel'dovich (tSZ) effect~\cite{Sunyaev:1970eu}.  These tSZ-identified clusters will be detected to high redshift, and will provide complementary information to optical measurements of the systems detected in both wavelength bands.  The amplitude of the tSZ effect for a given cluster provides a low-scatter proxy for the cluster mass, but the mapping of tSZ decrement to mass must be calibrated by additional observations. 
The synergy between CMB-S4 and LSST will be important for fully exploring the potential of tSZ-selected clusters as cosmological probes and extracting maximal information from both datasets.
For example, LSST weak lensing measurements of tSZ-identified galaxy clusters can provide a cluster mass calibration. 
They also will give an important sanity check on internal CMB-based lensing mass measurements, since the systematics affecting the two experiments are nearly independent.
Conversely, tSZ measurements and CMB lensing will provide additional methods for establishing masses for LSST-selected clusters.

Furthermore, since the tSZ effect is redshift-independent, SZ-selected clusters will span a large range in $z$. Cross-matching the cluster catalogs from CMB-S4 and LSST will provide high-precision photometric redshift estimates for tSZ clusters at $z\lesssim1.5$ and a lower redshift bound for even more distant tSZ-selected clusters.  Photometric redshift distributions in the directions of the tSZ-selected clusters may be sufficient to determine the redshifts for even higher-$z$ objects.

A mass- and redshift-calibrated cluster catalog produced by the combination of LSST and CMB-S4 will serve as a powerful probe of the growth of structure across cosmological times, and thus of the properties of dark energy and the sum of neutrino masses~\cite{Carlstrom:2002}.  The significant overlap between the survey regions of LSST and CMB-S4 ensures that this synergy will be possible across a large fraction of the sky.  

The combination of LSST and CMB-S4 data will also enable new analyses that will constrain the nature of dark energy using probes beyond galaxy clusters.  A variety of other cross-correlation based measurements, which can exploit information from CMB lensing and the kinetic and polarized Sunyaev-Zel'dovich effects as well as Rubin Observatory lensing and large-scale-structure measurements, will be enabled by the two experiments.  A separate white paper~\cite{Snowmass2021:JointProbes} describes these opportunities in more detail.

\section{Testing the Nature of Gravity with Wide-Area Spectroscopy}


Galaxies and weak lensing maps trace the same underlying matter field. Galaxies are biased tracers of matter but 
with spec-$z$'s can  provide full three-dimensional information, while
lensing maps are unbiased tracers that 
mainly provide information about projected density. Cross-correlating galaxies with lensing maps can provide very fine tomographic information about matter clustering independent of galaxy bias \citep[see, e.g.,][]{Baldauf2010,Mandelbaum2013}. 

As a result, combining cross-correlations between galaxy density and lensing with measurements of redshift-space distortions in galaxy clustering allows for tests of gravity on cosmological scales. Galaxy positions and velocities probe the behavior of non-relativistic matter under gravitation, while lensing is sensitive to the paths of purely relativistic particles (photons). For instance, previous work has defined a  statistic $E_G$ which combines information from these quantities and is sensitive to deviations from General Relativity \cite{Reyes2010, Zhang2007}. 
In order for the dependence on linear galaxy bias to cancel from $E_G$ (or similar statistics), redshift-space distortion measurements (which require spectroscopic redshift measurements) must be made on the same galaxy sample as that used to measure clustering and as lenses in galaxy-galaxy lensing analyses. Therefore, a sample of galaxies with spec-$z$'s that overlaps the LSST WL sample is required to measure $E_G$. 

DESI and 4MOST should provide such overlapping samples via both LRGs with $\bar{z} \approx 0.77$ and ELGs with $\bar{z}\approx 1.0$.  The combined samples of DESI and 4MOST ELGs and LRGs should each enable a determination of $E_G$ to $\sim 0.004$, roughly 10 times more precise than current constraints.  
Enlarging the northern LSST footprint (as proposed in Ref.~ \cite{BigSky} and discussed in the next section) would reduce errors by $\sim25\%$ more; greater improvements are possible with enlarged spectroscopic samples in the southern LSST footprint.

\section{Rubin Observatory Imaging to Enable a DESI-2 Survey}

\label{northern_lsst}

At the conclusion of the planned DESI survey (c. 2026), the instrument should still be a nearly-uniquely capable facility for wide-area spectroscopic surveys; it is unlikely to be surpassed by other projects until 2030 or later. The strategy for a follow-on, DESI-2 survey is still yet to be determined; however, the availability of $ugrizy$ imaging from Rubin Observatory should enable much more powerful target selection methods to be applied than were possible for the current DESI program.

A limiting factor, however, is that from its location in the Northern hemisphere, much of the LSST sky area can be observed only at high airmass (requiring long observation times) or not at all.  Many probes of cosmology, including Baryon Acoustic Oscillation measurements and tests for primordial non-Gaussianity from large-scale clustering, will be steadily more sensitive the wider area of sky that is covered.  As a result, we would wish to maximize the sky area that both has LSST-like $ugrizy$ imaging and is accessible to DESI.

The current LSST ``Wide, Fast, Deep'' main survey design covers the Northern sky up to declination of +12 degrees; most of the LSST footprint is far to the south.  However, due to its location at latitude +32$^{\circ}$~N, 
DESI can observe the far-Southern sky only inefficiently.  If DESI is restricted to declination $> -20^{\circ}$, the region where its survey speed would be degraded by at most $\sim 2\times$ due to having to point near the horizon, the DESI-LSST overlap would consist of roughly 8700 square degrees of sky.  This may be compared to the 14,000 square degrees which the current DESI survey is expected to cover.

In Ref. \cite{Capak2019} a mini-survey with LSST was previously recommended  which would enable this.  They proposed performing shallower observations with LSST covering declinations up to +30$^{\circ}$, primarily to improve photometric redshifts for the Euclid satellite.  Such data would also be sufficient to enable a DESI-2 survey of maximal sky area, \textbf{if the mini-survey observations are performed early in the LSST survey.}  Such front-loading is required because target selection for a second-generation DESI survey will need to be completed around 2026.  If such data could be obtained, it could greatly improve the targeting efficiency of a DESI-2 program, and hence its constraining power.  Extending the survey area to cover declinations from $-$20$^{\circ}$ to +30$^{\circ}$ would increase the DESI-2/LSST overlap to 13,300 square degrees, only a little short of the sky coverage of the current DESI survey. 






\section{Summary of Opportunities}

As described above, studies of dark energy via the Rubin Observatory can benefit greatly from additional data to follow up supernovae and strong lens systems, improve photometric redshifts, and constrain astrophysical systematics.  The facilities needed for such work can broadly be separated into those which enable photometry or spectroscopy of individual objects; those which provide wide-field, highly multiplexed spectroscopy; and those which provide high-multiplex spectroscopy over smaller fields of view on telescopes of much greater light-gathering power.  We summarize the opportunities to make gains via each of these classes of capabilities separately.

\subsection{Single-target imaging and spectroscopy} 

Cosmological studies based on SNe and strongly lensed systems in the new era of large-area, high-cadence optical surveys such as LSST will require single-object follow-up data of several types:

\begin{itemize}
    \item Spectrophotometry of SNe Ia will enable direct cosmological measurements, the development of spectral templates and the calibration of photometric classification performance at high redshift (where it is the most uncertain), as well as tests of systematic effects. Given the wide range of LSST SN brightnesses, this work will be reliant on large amounts of low-to-medium resolution spectroscopy on ground-based telescopes with apertures from 4~m to $>20$~m (and, ideally, space-based telescopes as well).
    \item Follow-up imaging with more frequent cadence than LSST will be valuable for measuring strong lensing time delays; this can generally be conducted on 2--4~m telescopes.
    \item Adaptive optics IFU spectroscopy, or AO imaging plus slit spectroscopy, is needed for precision image position measurements and lens system modeling; this work will require access to both 8--10~m and $>20$~m telescopes with AO capabilities.
\end{itemize}
The Kavli/NOAO/LSST report (summarized in Table~\ref{tab:resources}) provided quantitative estimates of the telescope time required to support cosmology measurements with LSST supernovae and strong-lens systems \cite{kavli}.

\begin{table}[htbp!]
    \centering
    \begin{tabular}{|c|c|}

        \hline
         Facility & Supernova single-object follow-up requirements  \\
         \hline
        4 m spectroscopy & 60--180 nights total \\
        8 m spectroscopy & 180--540 nights total \\
        $>20$m spectroscopy & 180--540 nights total \\ 
        \hline
        Facility & Strong Lensing single-object follow-up requirements  \\
        \hline
         2--4~m non-AO imaging & $<8000$ hours total \\
        $>8$m AO imaging & $\sim30$ hours, split amongst $>$8~m+ and $>20$~m telescopes\\
        $>8$m spectroscopy & $\sim100$ hours, split amongst $>$8~m+ and $>20$~m telescopes \\
         \hline
    \end{tabular}
    \caption{Summaries of the required resources, as estimated in the Kavli/NOAO/LSST study on Ground-based Optical/IR follow-up. \label{tab:resources}}
 
\end{table}

Although in many cases suitable instruments exist (e.g., Keck/OSIRIS) or are being developed (e.g., Gemini/SCORPIO), cases remain where the telescopes best positioned for this work have suboptimal instrumentation; as a result, additional development, not just telescope time, may be valuable.

\subsection{Wide-field spectroscopy}

LSST cosmological constraints can be significantly improved in a number of ways by incorporating additional wide-field, highly-multiplexed spectroscopy:
\begin{itemize}

\item Spectroscopic follow-up for strong lens systems and SNe/hosts will still be needed after 4MOST/TiDES observations are finished circa 2027 \cite{tides}.  This could be pursued via an extension of the 4MOST survey or by using PFS (for deep drilling field surveys), DESI, or a future spectroscopic facility.

\item Increasing the area in the southern hemisphere with DESI-like target selection of galaxies and quasars would improve cross-correlation calibration of photo-$z$'s and other cross-correlation science. This could be achieved by performing additional surveys with the 4MOST instrument covering thousands of square degrees.
  
\item Denser and/or higher-redshift sampling of galaxies over thousands of square degrees (e.g., in the Northern part of the LSST survey footprint) could significantly improve constraints on theories of modified gravity and models of intrinsic alignments for weak lensing studies. Such a project could be pursued efficiently with the DESI spectrograph, PFS at Subaru, or a future wide-field spectroscopic facility such as Megamapper \cite{megamapper}, the Maunakea Spectroscopic Explorer (MSE, \cite{mse}), or SpecTel \cite{spectel}. 
  
The optimal solution for wide-field spectroscopy spanning the full LSST footprint would be a highly-multiplexed, wide field-of-view instrument in the South, ideally on an aperture at least as large as LSST's; unfortunately no such capability exists in the US OIR system. Such a facility could improve LSST cosmology while simultaneously enabling new constraints of dark energy, as described in other Snowmass white papers \cite{Dawson_Snowmass,Ferraro_Snowmass}.  The MegaMapper, MSE, and SpecTel projects all would be well-suited for this role.

At the same time a facility of this type was targeting extragalactic objects for LSST cosmology and other dark energy studies, it could also pursue other projects with different fibers (e.g., wide-field surveys of stars in the Milky Way halo, which can constrain dark matter models; cf. \cite{Bechtol_Snowmass,Chakrabarti_Snowmass}.  This new capability should also lead to advances across a broad range of astrophysics, expanding the legacy of LSST and benefiting a diversity of communities. As a result, it is likely that such a facility could attract funding from multiple sources, building on the model of collaboration between HEP and other partners in the Sloan Digital Sky Survey and its successor programs.
\end{itemize}

\subsection{Deep but highly-multiplexed spectroscopy}

Access to modestly-wide-field, highly-multiplexed, large aperture spectroscopic facilities can lead to substantial gains in the constraining power of LSST and CMB-S4 compared to what would be possible without additional data:

\begin{itemize}

\item Improvements to deep photometric redshift training sets will yield better redshift estimates for individual objects, particularly improving constraints from large-scale structure and galaxy clusters.  

\item Tests of blending and intrinsic alignment effects will reduce the impact of critical systematics on cosmological measurements. 

\item Spectroscopy in a statistical sample of galaxy cluster fields will enable calibration of mass-observable relations and reductions in systematics.
\end{itemize}

For developing photometric redshift training sets as well as testing models of blending and intrinsic alignments, the key need is access to an instrument with maximal multiplexing with a field of view of at least 20$^\prime$ diameter on as large a telescope aperture as possible. It is likely that all of these goals can be pursued simultaneously with a common dataset. Subaru/PFS, Megamapper, MSE, SpecTel, and GMT/GMACS with the MANIFEST fiber feed are all well-suited for this work (wider-field fiber-fed spectrographs on other $>$6~m telescopes could also be suitable), as summarized in \autoref{table:photoz_times}.  
Personnel costs are likely to be high on smaller telescopes because of the longer time required for deep surveys on them, which may make instruments on 4~m telescopes such as DESI less attractive for this work.

The telescope time required to obtain data for photo-$z$ training, blending, and intrinsic alignment constraints is substantial, but it can be spread out over the ten-year span of the LSST survey, reducing the impact in any one year.  Additionally, these studies are highly synergistic with both the training needs for space-based experiments such as Euclid and the Nancy Grace Roman Space Telescope and with studies of galaxy evolution, potentially allowing combined surveys with a large impact on multiple fields of research and funded by multiple sources.

For galaxy cluster studies with both Rubin Observatory and CMB-S4, targeting more objects over smaller fields of view is desirable.  For that work, suitable options may include Gemini/GMACS, Keck/DEIMOS or LRIS, GMT/GMACS (in multislit mode), TMT/WFOS, or a new, higher-multiplex spectrograph at one of these observatories; wider-field facilities such as DESI, Subaru/PFS, Megamapper, MSE or SpecTel would be less well suited due to the lower fiber density and excessively large field of view.

\vskip 0.3in

In addition to the value that additional data can provide to LSST analyses, the Vera C. Rubin Observatory itself is uniquely capable of increasing the impact of a DESI-2 survey by providing modestly deep six-band imaging over a large area of sky.  A Northern mini-survey conducted in the first 1-2 years of LSST would enable optimized target selection methods to be employed uniformly over a DESI-2 footprint of more than 13,000 square degrees, which could provide great improvements to the capabilities of DESI-2 at low cost.

\vskip 0.3in

Above, we have discussed a wide variety of ways that smaller investments in other projects can enable improved dark energy constraints from upcoming, flagship dark energy experiments, allowing us to obtain the maximum impact from these flagship experiments. Obtaining this additional data in a timely fashion will both reduce statistical uncertainties (via expanded strong lens and supernova samples and photometric redshifts that extract the full potential out of LSST data, or via improved target selection and enlarged footprints for DESI-2 surveys) and can mitigate systematics which otherwise are expected to be comparable to statistical errors from LSST \cite{descsrd}. Many of these programs could be pursued at much smaller cost than a new Stage IV dark energy experiment, fitting well below the P5 small projects threshold, but offer similar improvements to constraining power as a separate new project.  For instance, comparatively modest amounts of funding could enable the purchase of dedicated telescope time, in-kind contributions of new instrumentation or components with guaranteed time provided in compensation, or the collection and analysis of datasets obtained via telescope time obtained via either proposal processes or international contributions.

In contrast, a new dedicated spectroscopic facility would require major investment.  In the short term, funding to enable instrument improvements and develop the international collaborations needed to make such a project happen, with a construction start before the end of this decade if possible, would be valuable. Such a new capability would not only provide new constraints on cosmology of its own as described in other white papers, but also would be capable of improving the science yield from the LSST imaging survey, enabling us to obtain the maximum yield from this unprecedented dataset. It may not be possible to obtain the deep spectroscopic datasets that can enable improved photometric redshift training and the characterization of blending and intrinsic alignments before such a facility is available, unless dedicated funding or in-kind contributions for such observations are made available. In that case, a new spectroscopic telescope could enable improved constraints on cosmology for the final LSST analyses in the middle of the next decade, as well as pursuing a variety of explorations of the dark sector of its own that would utilize LSST data for target selection.  

\newpage
\begin{center}
{\large\bf Acknowledgements}    
    
\end{center}
Members of the LSST Dark Energy Science Collaboration acknowledge ongoing support from the Institut National de Physique Nucl\'eaire et de Physique des Particules in France; the Science \& Technology Facilities Council in the United Kingdom; and the Department of Energy, the National Science Foundation, and the LSST Corporation in the United States.  DESC uses resources of the IN2P3 Computing Center (CC-IN2P3--Lyon/Villeurbanne - France) funded by the Centre National de la Recherche Scientifique; the National Energy Research Scientific Computing Center, a DOE Office of Science User Facility supported by the Office of Science of the U.S.\ Department of Energy under Contract No.\ DE-AC02-05CH11231; STFC DiRAC HPC Facilities, funded by UK BIS National E-infrastructure capital grants; and the UK particle physics grid, supported by the GridPP Collaboration.  This work was performed in part under DOE Contract DE-AC02-76SF00515.



\bibliographystyle{unsrt}
\bibliography{main.bib}

\begin{thebibliography}{100}

\bibitem{descsrd}
{The LSST Dark Energy Science Collaboration}, R.~{Mandelbaum}, T.~{Eifler},
  R.~{Hlo{\v z}ek}, T.~{Collett}, E.~{Gawiser}, D.~{Scolnic}, D.~{Alonso},
  H.~{Awan}, R.~{Biswas}, J.~{Blazek}, P.~{Burchat}, N.~E. {Chisari},
  I.~{Dell'Antonio}, S.~{Digel}, J.~{Frieman}, D.~A. {Goldstein}, I.~{Hook},
  {\v Z}.~{Ivezi{\'c}}, S.~M. {Kahn}, S.~{Kamath}, D.~{Kirkby}, T.~{Kitching},
  E.~{Krause}, P.-F. {Leget}, P.~J. {Marshall}, J.~{Meyers}, H.~{Miyatake},
  J.~A. {Newman}, R.~{Nichol}, E.~{Rykoff}, F.~J. {Sanchez}, A.~{Slosar},
  M.~{Sullivan}, and M.~A. {Troxel}.
\newblock {The LSST Dark Energy Science Collaboration (DESC) Science
  Requirements Document}.
\newblock {\em arXiv:1809.01669}, September 2018.

\bibitem{Snowmass2021:JointProbes}
E.~Baxter, C.~Chang, A.~Hearin, et~al.
\newblock {Snowmass2021 Cosmic Frontier White Paper: Opportunities from
  Cross-survey Analyses of Static Probes}.
\newblock \emph{Snowmass 2021 White Paper}, 2022.

\bibitem{Snowmass2013:Transient}
Alex~G. {Kim}, Antonella {Palmese}, Maria E.~S. {Pereira}, Greg {Aldering},
  Felipe {Andrade-Oliveira}, James {Annis}, Stephen {Bailey}, Segev {BenZvi},
  Ulysses {Braga-Neto}, Fr{\'e}d{\'e}ric {Courbin}, Alyssa {Garcia}, David
  {Jeffery}, Gautham {Narayan}, Saul {Perlmutter}, Marcelle {Soares-Santos},
  Tommaso {Treu}, and Lifan {Wang}.
\newblock {Snowmass2021 Cosmic Frontier CF6 White Paper: Multi-Experiment
  Probes for Dark Energy -- Transients}.
\newblock {\em arXiv e-prints}, page arXiv:2203.11226, March 2022.

\bibitem{ztfnat}
E.~{Bellm} and S.~{Kulkarni}.
\newblock {The unblinking eye on the sky}.
\newblock {\em Nature Astronomy}, 1:0071, March 2017.

\bibitem{LOI-LS4}
P.~E. Nugent et~al.
\newblock {La Silla Schmidt Southern Survey}.
\newblock Snowmass 2020 LOI, 2020.

\bibitem{deJaeger2017}
T.~{de Jaeger}, L.~{Galbany}, A.~V. {Filippenko}, S.~{Gonz{\'a}lez-Gait{\'a}n},
  N.~{Yasuda}, K.~{Maeda}, M.~{Tanaka}, T.~{Morokuma}, T.~J. {Moriya},
  N.~{Tominaga}, K.~{Nomoto}, Y.~{Komiyama}, J.~P. {Anderson}, T.~G. {Brink},
  R.~G. {Carlberg}, G.~{Folatelli}, M.~{Hamuy}, G.~{Pignata}, and W.~{Zheng}.
\newblock {SN 2016jhj at redshift 0.34: extending the Type II supernova Hubble
  diagram using the standard candle method}.
\newblock {\em \mnras}, 472:4233--4243, Dec 2017.

\bibitem{inserra2018}
C.~{Inserra}, R.~C. {Nichol}, D.~{Scovacricchi}, J.~{Amiaux}, M.~{Brescia},
  C.~{Burigana}, E.~{Cappellaro}, C.~S. {Carvalho}, S.~{Cavuoti},
  V.~{Conforti}, J.~C. {Cuilland re}, A.~{da Silva}, A.~{De Rosa}, M.~{Della
  Valle}, J.~{Dinis}, E.~{Franceschi}, I.~{Hook}, P.~{Hudelot}, K.~{Jahnke},
  T.~{Kitching}, H.~{Kurki-Suonio}, I.~{Lloro}, G.~{Longo}, E.~{Maiorano},
  M.~{Maris}, J.~D. {Rhodes}, R.~{Scaramella}, S.~J. {Smartt}, M.~{Sullivan},
  C.~{Tao}, R.~{Toledo-Moreo}, I.~{Tereno}, M.~{Trifoglio}, and
  L.~{Valenziano}.
\newblock {Euclid: Superluminous supernovae in the Deep Survey}.
\newblock {\em \aap}, 609:A83, Jan 2018.

\bibitem{Saunders2018}
C.~{Saunders}, G.~{Aldering}, P.~{Antilogus}, S.~{Bailey}, C.~{Baltay},
  K.~{Barbary}, D.~{Baugh}, K.~{Boone}, S.~{Bongard}, C.~{Buton}, J.~{Chen},
  N.~{Chotard}, Y.~{Copin}, S.~{Dixon}, P.~{Fagrelius}, H.~K. {Fakhouri},
  U.~{Feindt}, D.~{Fouchez}, E.~{Gangler}, B.~{Hayden}, W.~{Hillebrandt}, A.~G.
  {Kim}, M.~{Kowalski}, D.~{K{\"u}sters}, P.~F. {Leget}, S.~{Lombardo},
  J.~{Nordin}, R.~{Pain}, E.~{Pecontal}, R.~{Pereira}, S.~{Perlmutter},
  D.~{Rabinowitz}, M.~{Rigault}, D.~{Rubin}, K.~{Runge}, G.~{Smadja},
  C.~{Sofiatti}, N.~{Suzuki}, C.~{Tao}, S.~{Taubenberger}, R.~C. {Thomas},
  M.~{Vincenzi}, and The {Nearby Supernova Factory}.
\newblock {SNEMO: Improved Empirical Models for Type Ia Supernovae}.
\newblock {\em \apj}, 869:167, Dec 2018.

\bibitem{Fakhouri2015}
H.~K. {Fakhouri}, K.~{Boone}, G.~{Aldering}, P.~{Antilogus}, C.~{Aragon},
  S.~{Bailey}, C.~{Baltay}, K.~{Barbary}, D.~{Baugh}, S.~{Bongard}, C.~{Buton},
  J.~{Chen}, M.~{Childress}, N.~{Chotard}, Y.~{Copin}, P.~{Fagrelius},
  U.~{Feindt}, M.~{Fleury}, D.~{Fouchez}, E.~{Gangler}, B.~{Hayden}, A.~G.
  {Kim}, M.~{Kowalski}, P.~F. {Leget}, S.~{Lombardo}, J.~{Nordin}, R.~{Pain},
  E.~{Pecontal}, R.~{Pereira}, S.~{Perlmutter}, D.~{Rabinowitz}, J.~{Ren},
  M.~{Rigault}, D.~{Rubin}, K.~{Runge}, C.~{Saunders}, R.~{Scalzo},
  G.~{Smadja}, C.~{Sofiatti}, M.~{Strovink}, N.~{Suzuki}, C.~{Tao}, R.~C.
  {Thomas}, B.~A. {Weaver}, and The {Nearby Supernova Factory}.
\newblock {Improving Cosmological Distance Measurements Using Twin Type Ia
  Supernovae}.
\newblock {\em \apj}, 815:58, Dec 2015.

\bibitem{2018arXiv181102374D}
{DES Collaboration}, T.~M.~C. {Abbott}, S.~{Allam}, P.~{Andersen}, C.~{Angus},
  J.~{Asorey}, A.~{Avelino}, S.~{Avila}, B.~A. {Bassett}, K.~{Bechtol}, G.~M.
  {Bernstein}, E.~{Bertin}, D.~{Brooks}, D.~{Brout}, P.~{Brown}, D.~L. {Burke},
  J.~{Calcino}, A.~{Carnero Rosell}, D.~{Carollo}, M.~{Carrasco Kind},
  J.~{Carretero}, R.~{Casas}, F.~J. {Castander}, R.~{Cawthon}, P.~{Challis},
  M.~{Childress}, A.~{Clocchiatti}, C.~E. {Cunha}, C.~B. {D'Andrea}, L.~N. {da
  Costa}, C.~{Davis}, T.~M. {Davis}, J.~{De Vicente}, D.~L. {DePoy},
  S.~{Desai}, H.~T. {Diehl}, P.~{Doel}, A.~{Drlica-Wagner}, T.~F. {Eifler},
  A.~E. {Evrard}, E.~{Fernandez}, A.~V. {Filippenko}, D.~A. {Finley},
  B.~{Flaugher}, R.~J. {Foley}, P.~{Fosalba}, J.~{Frieman}, L.~{Galbany},
  J.~{Garcia-Bellido}, E.~{Gaztanaga}, T.~{Giannantonio}, K.~{Glazebrook},
  D.~A. {Goldstein}, S.~{Gonzalez-Gaitan}, D.~{Gruen}, R.~A. {Gruendl},
  J.~{Gschwend}, R.~R. {Gupta}, G.~{Gutierrez}, W.~G. {Hartley}, S.~R.
  {Hinton}, D.~L. {Hollowood}, K.~{Honscheid}, J.~K. {Hoormann}, B.~{Hoyle},
  D.~J. {James}, T.~{Jeltema}, M.~W.~G. {Johnson}, M.~D. {Johnson}, E.~{Kasai},
  S.~{Kent}, R.~{Kessler}, A.~G. {Kim}, R.~P. {Kirshner}, E.~{Kovacs},
  E.~{Krause}, R.~{Kron}, K.~{Kuehn}, S.~{Kuhlmann}, N.~{Kuropatkin},
  O.~{Lahav}, J.~{Lasker}, G.~F. {Lewis}, T.~S. {Li}, C.~{Lidman}, M.~{Lima},
  H.~{Lin}, E.~{Macaulay}, M.~A.~G. {Maia}, K.~S. {Mandel}, M.~{March},
  J.~{Marriner}, J.~L. {Marshall}, P.~{Martini}, F.~{Menanteau}, C.~J.
  {Miller}, R.~{Miquel}, V.~{Miranda}, J.~J. {Mohr}, E.~{Morganson},
  D.~{Muthukrishna}, A.~{M{\"o}ller}, E.~{Neilsen}, R.~C. {Nichol}, B.~{Nord},
  P.~{Nugent}, R.~L.~C. {Ogando}, A.~{Palmese}, Y.~C. {Pan}, A.~A. {Plazas},
  M.~{Pursiainen}, A.~K. {Romer}, A.~{Roodman}, E.~{Rozo}, E.~S. {Rykoff},
  M.~{Sako}, E.~{Sanchez}, V.~{Scarpine}, R.~{Schindler}, M.~{Schubnell},
  D.~{Scolnic}, S.~{Serrano}, I.~{Sevilla-Noarbe}, R.~{Sharp}, M.~{Smith},
  M.~{Soares-Santos}, F.~{Sobreira}, N.~E. {Sommer}, H.~{Spinka}, E.~{Suchyta},
  M.~{Sullivan}, E.~{Swann}, G.~{Tarle}, D.~{Thomas}, R.~C. {Thomas}, M.~A.
  {Troxel}, B.~E. {Tucker}, S.~A. {Uddin}, A.~R. {Walker}, W.~{Wester},
  P.~{Wiseman}, R.~C. {Wolf}, B.~{Yanny}, B.~{Zhang}, and Y.~{Zhang}.
\newblock {First Cosmology Results using Type Ia Supernovae from the Dark
  Energy Survey: Constraints on Cosmological Parameters}.
\newblock {\em arXiv e-prints}, page arXiv:1811.02374, Nov 2018.

\bibitem{2016ApJS..225...31L}
M.~{Lochner}, J.~D. {McEwen}, H.~V. {Peiris}, O.~{Lahav}, and M.~K. {Winter}.
\newblock {Photometric Supernova Classification with Machine Learning}.
\newblock {\em \apjs}, 225:31, August 2016.

\bibitem{hlozek2012}
Ren{\'e}e {Hlozek}, Martin {Kunz}, Bruce {Bassett}, Mat {Smith}, James
  {Newling}, Melvin {Varughese}, Rick {Kessler}, Joseph~P. {Bernstein}, Heather
  {Campbell}, Ben {Dilday}, Bridget {Falck}, Joshua {Frieman}, Steve
  {Kuhlmann}, Hubert {Lampeitl}, John {Marriner}, Robert~C. {Nichol}, Adam~G.
  {Riess}, Masao {Sako}, and Donald~P. {Schneider}.
\newblock {Photometric Supernova Cosmology with BEAMS and SDSS-II}.
\newblock {\em \apj}, 752:79, Jun 2012.

\bibitem{campbell2013}
Heather {Campbell}, Chris~B. {D'Andrea}, Robert~C. {Nichol}, Masao {Sako},
  Mathew {Smith}, Hubert {Lampeitl}, Matthew~D. {Olmstead}, Bruce {Bassett},
  Rahul {Biswas}, Peter {Brown}, David {Cinabro}, Kyle~S. {Dawson}, Ben
  {Dilday}, Ryan~J. {Foley}, Joshua~A. {Frieman}, Peter {Garnavich}, Renee
  {Hlozek}, Saurabh~W. {Jha}, Steve {Kuhlmann}, Martin {Kunz}, John {Marriner},
  Ramon {Miquel}, Michael {Richmond}, Adam {Riess}, Donald~P. {Schneider},
  Jesper {Sollerman}, Matt {Taylor}, and Gong-Bo {Zhao}.
\newblock {Cosmology with Photometrically Classified Type Ia Supernovae from
  the SDSS-II Supernova Survey}.
\newblock {\em \apj}, 763:88, Feb 2013.

\bibitem{jones2018}
D.~O. {Jones}, D.~M. {Scolnic}, A.~G. {Riess}, A.~{Rest}, R.~P. {Kirshner},
  E.~{Berger}, R.~{Kessler}, Y.~C. {Pan}, R.~J. {Foley}, R.~{Chornock}, C.~A.
  {Ortega}, P.~J. {Challis}, W.~S. {Burgett}, K.~C. {Chambers}, P.~W. {Draper},
  H.~{Flewelling}, M.~E. {Huber}, N.~{Kaiser}, R.~P. {Kudritzki},
  N.~{Metcalfe}, J.~{Tonry}, R.~J. {Wainscoat}, C.~{Waters}, E.~E.~E. {Gall},
  R.~{Kotak}, M.~{McCrum}, S.~J. {Smartt}, and K.~W. {Smith}.
\newblock {Measuring Dark Energy Properties with Photometrically Classified
  Pan-STARRS Supernovae. II. Cosmological Parameters}.
\newblock {\em \apj}, 857:51, Apr 2018.

\bibitem{kavli}
J.~{Najita}, B.~{Willman}, D.~P. {Finkbeiner}, R.~J. {Foley}, S.~{Hawley},
  J.~A. {Newman}, G.~{Rudnick}, J.~D. {Simon}, D.~{Trilling}, R.~{Street},
  A.~{Bolton}, R.~{Angus}, E.~F. {Bell}, D.~{Buzasi}, D.~{Ciardi}, J.~R.~A.
  {Davenport}, W.~{Dawson}, M.~{Dickinson}, A.~{Drlica-Wagner}, J.~{Elias},
  D.~{Erb}, L.~{Feaga}, W.-f. {Fong}, E.~{Gawiser}, M.~{Giampapa},
  P.~{Guhathakurta}, J.~L. {Hoffman}, H.~{Hsieh}, E.~{Jennings}, K.~V.
  {Johnston}, V.~{Kashyap}, T.~S. {Li}, E.~{Linder}, R.~{Mandelbaum},
  P.~{Marshall}, T.~{Matheson}, S.~{Meibom}, B.~W. {Miller}, J.~{O'Meara},
  V.~{Reddy}, S.~{Ridgway}, C.~M. {Rockosi}, D.~J. {Sand}, C.~{Schafer},
  S.~{Schmidt}, B.~{Sesar}, S.~S. {Sheppard}, C.~A. {Thomas}, E.~J. {Tollerud},
  J.~{Trump}, and A.~{von der Linden}.
\newblock {Maximizing Science in the Era of LSST: A Community-Based Study of
  Needed US Capabilities}.
\newblock {\em arXiv e-prints}, page arXiv:1610.01661, October 2016.

\bibitem{ozdes}
M.~J. {Childress}, C.~{Lidman}, T.~M. {Davis}, B.~E. {Tucker}, J.~{Asorey},
  F.~{Yuan}, T.~M.~C. {Abbott}, F.~B. {Abdalla}, S.~{Allam}, J.~{Annis},
  M.~{Banerji}, A.~{Benoit-L{\'e}vy}, S.~R. {Bernard}, E.~{Bertin},
  D.~{Brooks}, E.~{Buckley-Geer}, D.~L. {Burke}, A.~{Carnero Rosell},
  D.~{Carollo}, M.~{Carrasco Kind}, J.~{Carretero}, F.~J. {Castander}, C.~E.
  {Cunha}, L.~N. {da Costa}, C.~B. {D'Andrea}, P.~{Doel}, T.~F. {Eifler}, A.~E.
  {Evrard}, B.~{Flaugher}, R.~J. {Foley}, P.~{Fosalba}, J.~{Frieman},
  J.~{Garc{\'{\i}}a-Bellido}, K.~{Glazebrook}, D.~A. {Goldstein}, D.~{Gruen},
  R.~A. {Gruendl}, J.~{Gschwend}, R.~R. {Gupta}, G.~{Gutierrez}, S.~R.
  {Hinton}, J.~K. {Hoormann}, D.~J. {James}, R.~{Kessler}, A.~G. {Kim}, A.~L.
  {King}, E.~{Kovacs}, K.~{Kuehn}, S.~{Kuhlmann}, N.~{Kuropatkin}, D.~J.
  {Lagattuta}, G.~F. {Lewis}, T.~S. {Li}, M.~{Lima}, H.~{Lin}, E.~{Macaulay},
  M.~A.~G. {Maia}, J.~{Marriner}, M.~{March}, J.~L. {Marshall}, P.~{Martini},
  R.~G. {McMahon}, F.~{Menanteau}, R.~{Miquel}, A.~{Moller}, E.~{Morganson},
  J.~{Mould}, D.~{Mudd}, D.~{Muthukrishna}, R.~C. {Nichol}, B.~{Nord}, R.~L.~C.
  {Ogando}, F.~{Ostrovski}, D.~{Parkinson}, A.~A. {Plazas}, S.~L. {Reed},
  K.~{Reil}, A.~K. {Romer}, E.~S. {Rykoff}, M.~{Sako}, E.~{Sanchez},
  V.~{Scarpine}, R.~{Schindler}, M.~{Schubnell}, D.~{Scolnic},
  I.~{Sevilla-Noarbe}, N.~{Seymour}, R.~{Sharp}, M.~{Smith},
  M.~{Soares-Santos}, F.~{Sobreira}, N.~E. {Sommer}, H.~{Spinka}, E.~{Suchyta},
  M.~{Sullivan}, M.~E.~C. {Swanson}, G.~{Tarle}, S.~A. {Uddin}, A.~R. {Walker},
  W.~{Wester}, and B.~R. {Zhang}.
\newblock {OzDES multifibre spectroscopy for the Dark Energy Survey: 3-yr
  results and first data release}.
\newblock {\em \mnras}, 472:273--288, November 2017.

\bibitem{tides}
E.~{Swann}, M.~{Sullivan}, J.~{Carrick}, S.~{Hoenig}, I.~{Hook}, R.~{Kotak},
  K.~{Maguire}, R.~{Nichol}, and S.~{Smartt}.
\newblock {4MOST Consortium Survey 10: The Time-Domain Extragalactic Survey
  (TiDES)}.
\newblock {\em arXiv e-prints}, page arXiv:1903.02476, Mar 2019.

\bibitem{Chakrabarti2022}
Sukanya Chakrabarti et~al.
\newblock {Snowmass2021 Cosmic Frontier White Paper: Observational Facilities
  to Study Dark Matter}.
\newblock 3 2022.

\bibitem{specneeds}
J.~A. {Newman}, A.~{Abate}, F.~B. {Abdalla}, S.~{Allam}, S.~W. {Allen},
  R.~{Ansari}, S.~{Bailey}, W.~A. {Barkhouse}, T.~C. {Beers}, M.~R. {Blanton},
  M.~{Brodwin}, J.~R. {Brownstein}, R.~J. {Brunner}, M.~{Carrasco Kind}, J.~L.
  {Cervantes-Cota}, E.~{Cheu}, N.~E. {Chisari}, M.~{Colless}, J.~{Comparat},
  J.~{Coupon}, C.~E. {Cunha}, A.~{de la Macorra}, I.~P. {Dell'Antonio}, B.~L.
  {Frye}, E.~J. {Gawiser}, N.~{Gehrels}, K.~{Grady}, A.~{Hagen}, P.~B. {Hall},
  A.~P. {Hearin}, H.~{Hildebrandt}, C.~M. {Hirata}, S.~{Ho}, K.~{Honscheid},
  D.~{Huterer}, {\v Z}.~{Ivezi{\'c}}, J.-P. {Kneib}, J.~W. {Kruk}, O.~{Lahav},
  R.~{Mandelbaum}, J.~L. {Marshall}, D.~J. {Matthews}, B.~{M{\'e}nard},
  R.~{Miquel}, M.~{Moniez}, H.~W. {Moos}, J.~{Moustakas}, A.~D. {Myers},
  C.~{Papovich}, J.~A. {Peacock}, C.~{Park}, M.~{Rahman}, J.~{Rhodes}, J.-S.
  {Ricol}, I.~{Sadeh}, A.~{Slozar}, S.~J. {Schmidt}, D.~K. {Stern}, J.~{Anthony
  Tyson}, A.~{von der Linden}, R.~H. {Wechsler}, W.~M. {Wood-Vasey}, and A.~R.
  {Zentner}.
\newblock {Spectroscopic needs for imaging dark energy experiments}.
\newblock {\em Astroparticle Physics}, 63:81--100, March 2015.

\bibitem{refsdal64a}
S.~{Refsdal}.
\newblock {The gravitational lens effect}.
\newblock {\em \mnras}, 128:295, 1964.

\bibitem{blandfordandnarayan92}
R.~D. {Blandford} and R.~{Narayan}.
\newblock {Cosmological applications of gravitational lensing}.
\newblock {\em \araa}, 30:311--358, 1992.

\bibitem{kelly}
P.~L. {Kelly}, S.~A. {Rodney}, T.~{Treu}, R.~J. {Foley}, G.~{Brammer}, K.~B.
  {Schmidt}, A.~{Zitrin}, A.~{Sonnenfeld}, L.-G. {Strolger}, O.~{Graur}, A.~V.
  {Filippenko}, S.~W. {Jha}, A.~G. {Riess}, M.~{Bradac}, B.~J. {Weiner},
  D.~{Scolnic}, M.~A. {Malkan}, A.~{von der Linden}, M.~{Trenti}, J.~{Hjorth},
  R.~{Gavazzi}, A.~{Fontana}, J.~C. {Merten}, C.~{McCully}, T.~{Jones},
  M.~{Postman}, A.~{Dressler}, B.~{Patel}, S.~B. {Cenko}, M.~L. {Graham}, and
  B.~E. {Tucker}.
\newblock {Multiple images of a highly magnified supernova formed by an
  early-type cluster galaxy lens}.
\newblock {\em Science}, 347:1123--1126, March 2015.

\bibitem{goobar16}
A.~{Goobar}, R.~{Amanullah}, S.~R. {Kulkarni}, P.~E. {Nugent}, J.~{Johansson},
  C.~{Steidel}, D.~{Law}, E.~{M{\"o}rtsell}, R.~{Quimby}, N.~{Blagorodnova},
  A.~{Brandeker}, Y.~{Cao}, A.~{Cooray}, R.~{Ferretti}, C.~{Fremling},
  L.~{Hangard}, M.~{Kasliwal}, T.~{Kupfer}, R.~{Lunnan}, F.~{Masci}, A.~A.
  {Miller}, H.~{Nayyeri}, J.~D. {Neill}, E.~O. {Ofek}, S.~{Papadogiannakis},
  T.~{Petrushevska}, V.~{Ravi}, J.~{Sollerman}, M.~{Sullivan}, F.~{Taddia},
  R.~{Walters}, D.~{Wilson}, L.~{Yan}, and O.~{Yaron}.
\newblock {iPTF16geu: A multiply imaged, gravitationally lensed type Ia
  supernova}.
\newblock {\em Science}, 356:291--295, April 2017.

\bibitem{gnkc17}
D.~A. {Goldstein}, P.~E. {Nugent}, D.~N. {Kasen}, and T.~E. {Collett}.
\newblock {Precise Time Delays from Chromatically Microlensed Type Ia
  Supernovae}.
\newblock {\em ArXiv e-prints}, page arXiv:1708.00003, July 2017.

\bibitem{riess16}
A.~G. {Riess}, L.~M. {Macri}, S.~L. {Hoffmann}, D.~{Scolnic}, S.~{Casertano},
  A.~V. {Filippenko}, B.~E. {Tucker}, M.~J. {Reid}, D.~O. {Jones}, J.~M.
  {Silverman}, R.~{Chornock}, P.~{Challis}, W.~{Yuan}, P.~J. {Brown}, and R.~J.
  {Foley}.
\newblock {A 2.4\% Determination of the Local Value of the Hubble Constant}.
\newblock {\em \apj}, 826:56, July 2016.

\bibitem{planckhitens}
{Planck Collaboration}, N.~{Aghanim}, M.~{Ashdown}, J.~{Aumont},
  C.~{Baccigalupi}, M.~{Ballardini}, A.~J. {Banday}, R.~B. {Barreiro},
  N.~{Bartolo}, S.~{Basak}, R.~{Battye}, K.~{Benabed}, J.-P. {Bernard},
  M.~{Bersanelli}, P.~{Bielewicz}, J.~J. {Bock}, A.~{Bonaldi}, L.~{Bonavera},
  J.~R. {Bond}, J.~{Borrill}, F.~R. {Bouchet}, F.~{Boulanger}, M.~{Bucher},
  C.~{Burigana}, R.~C. {Butler}, E.~{Calabrese}, J.-F. {Cardoso}, J.~{Carron},
  A.~{Challinor}, H.~C. {Chiang}, L.~P.~L. {Colombo}, C.~{Combet}, B.~{Comis},
  A.~{Coulais}, B.~P. {Crill}, A.~{Curto}, F.~{Cuttaia}, R.~J. {Davis}, P.~{de
  Bernardis}, A.~{de Rosa}, G.~{de Zotti}, J.~{Delabrouille}, J.-M. {Delouis},
  E.~{Di Valentino}, C.~{Dickinson}, J.~M. {Diego}, O.~{Dor{\'e}},
  M.~{Douspis}, A.~{Ducout}, X.~{Dupac}, G.~{Efstathiou}, F.~{Elsner}, T.~A.
  {En{\ss}lin}, H.~K. {Eriksen}, E.~{Falgarone}, Y.~{Fantaye}, F.~{Finelli},
  F.~{Forastieri}, M.~{Frailis}, A.~A. {Fraisse}, E.~{Franceschi}, A.~{Frolov},
  S.~{Galeotta}, S.~{Galli}, K.~{Ganga}, R.~T. {G{\'e}nova-Santos},
  M.~{Gerbino}, T.~{Ghosh}, J.~{Gonz{\'a}lez-Nuevo}, K.~M. {G{\'o}rski},
  S.~{Gratton}, A.~{Gruppuso}, J.~E. {Gudmundsson}, F.~K. {Hansen}, G.~{Helou},
  S.~{Henrot-Versill{\'e}}, D.~{Herranz}, E.~{Hivon}, Z.~{Huang},
  S.~{Ili{\'c}}, A.~H. {Jaffe}, W.~C. {Jones}, E.~{Keih{\"a}nen},
  R.~{Keskitalo}, T.~S. {Kisner}, L.~{Knox}, N.~{Krachmalnicoff}, M.~{Kunz},
  H.~{Kurki-Suonio}, G.~{Lagache}, J.-M. {Lamarre}, M.~{Langer}, A.~{Lasenby},
  M.~{Lattanzi}, C.~R. {Lawrence}, M.~{Le Jeune}, J.~P. {Leahy}, F.~{Levrier},
  M.~{Liguori}, P.~B. {Lilje}, M.~{L{\'o}pez-Caniego}, Y.-Z. {Ma}, J.~F.
  {Mac{\'{\i}}as-P{\'e}rez}, G.~{Maggio}, A.~{Mangilli}, M.~{Maris}, P.~G.
  {Martin}, E.~{Mart{\'{\i}}nez-Gonz{\'a}lez}, S.~{Matarrese}, N.~{Mauri},
  J.~D. {McEwen}, P.~R. {Meinhold}, A.~{Melchiorri}, A.~{Mennella},
  M.~{Migliaccio}, M.-A. {Miville-Desch{\^e}nes}, D.~{Molinari}, A.~{Moneti},
  L.~{Montier}, G.~{Morgante}, A.~{Moss}, S.~{Mottet}, P.~{Naselsky},
  P.~{Natoli}, C.~A. {Oxborrow}, L.~{Pagano}, D.~{Paoletti}, B.~{Partridge},
  G.~{Patanchon}, L.~{Patrizii}, O.~{Perdereau}, L.~{Perotto}, V.~{Pettorino},
  F.~{Piacentini}, S.~{Plaszczynski}, L.~{Polastri}, G.~{Polenta}, J.-L.
  {Puget}, J.~P. {Rachen}, B.~{Racine}, M.~{Reinecke}, M.~{Remazeilles},
  A.~{Renzi}, G.~{Rocha}, M.~{Rossetti}, G.~{Roudier}, J.~A.
  {Rubi{\~n}o-Mart{\'{\i}}n}, B.~{Ruiz-Granados}, L.~{Salvati}, M.~{Sandri},
  M.~{Savelainen}, D.~{Scott}, G.~{Sirri}, R.~{Sunyaev}, A.-S. {Suur-Uski},
  J.~A. {Tauber}, M.~{Tenti}, L.~{Toffolatti}, M.~{Tomasi}, M.~{Tristram},
  T.~{Trombetti}, J.~{Valiviita}, F.~{Van Tent}, L.~{Vibert}, P.~{Vielva},
  F.~{Villa}, N.~{Vittorio}, B.~D. {Wandelt}, R.~{Watson}, I.~K. {Wehus},
  M.~{White}, A.~{Zacchei}, and A.~{Zonca}.
\newblock {Planck intermediate results. XLVI. Reduction of large-scale
  systematic effects in HFI polarization maps and estimation of the
  reionization optical depth}.
\newblock {\em \aap}, 596:A107, December 2016.

\bibitem{phantom}
E.~{Di Valentino}, A.~{Melchiorri}, and J.~{Silk}.
\newblock {Reconciling Planck with the local value of H$_{0}$ in extended
  parameter space}.
\newblock {\em Physics Letters B}, 761:242--246, October 2016.

\bibitem{freedman17}
W.~L. {Freedman}.
\newblock {Cosmology at a crossroads}.
\newblock {\em Nature Astronomy}, 1:0121, May 2017.

\bibitem{zhao17}
G.-B. {Zhao}, M.~{Raveri}, L.~{Pogosian}, Y.~{Wang}, R.~G. {Crittenden}, W.~J.
  {Handley}, W.~J. {Percival}, F.~{Beutler}, J.~{Brinkmann}, C.-H. {Chuang},
  A.~J. {Cuesta}, D.~J. {Eisenstein}, F.-S. {Kitaura}, K.~{Koyama},
  B.~{L'Huillier}, R.~C. {Nichol}, M.~M. {Pieri}, S.~{Rodriguez-Torres}, A.~J.
  {Ross}, G.~{Rossi}, A.~G. {S{\'a}nchez}, A.~{Shafieloo}, J.~L. {Tinker},
  R.~{Tojeiro}, J.~A. {Vazquez}, and H.~{Zhang}.
\newblock {Dynamical dark energy in light of the latest observations}.
\newblock {\em Nature Astronomy}, 1:627--632, September 2017.

\bibitem{efstathiou14}
G.~{Efstathiou}.
\newblock {H$_{0}$ revisited}.
\newblock {\em \mnras}, 440:1138--1152, May 2014.

\bibitem{trouble}
J.~L. {Bernal}, L.~{Verde}, and A.~G. {Riess}.
\newblock {The trouble with H$_{0}$}.
\newblock {\em JCAP}, 10:019, October 2016.

\bibitem{kochanek02}
C.~S. {Kochanek}.
\newblock {What Do Gravitational Lens Time Delays Measure?}
\newblock {\em \apj}, 578:25--32, October 2002.

\bibitem{treu2010}
T.~{Treu}.
\newblock {Strong Lensing by Galaxies}.
\newblock {\em \araa}, 48:87--125, September 2010.

\bibitem{fassnacht11}
C.~D. {Fassnacht}, L.~V.~E. {Koopmans}, and K.~C. {Wong}.
\newblock {Galaxy number counts and implications for strong lensing}.
\newblock {\em \mnras}, 410:2167--2179, February 2011.

\bibitem{cosmography}
T.~{Treu} and P.~J. {Marshall}.
\newblock {Time delay cosmography}.
\newblock {\em AAPR}, 24:11, July 2016.

\bibitem{refsdal64b}
S.~{Refsdal}.
\newblock {On the possibility of determining Hubble's parameter and the masses
  of galaxies from the gravitational lens effect}.
\newblock {\em \mnras}, 128:307, 1964.

\bibitem{oguri03}
M.~{Oguri} and Y.~{Kawano}.
\newblock {Gravitational lens time delays for distant supernovae: breaking the
  degeneracy between radial mass profiles and the Hubble constant}.
\newblock {\em \mnras}, 338:L25--L29, February 2003.

\bibitem{collettcunnington16}
T.~E. {Collett} and S.~D. {Cunnington}.
\newblock {Observational selection biases in time-delay strong lensing and
  their impact on cosmography}.
\newblock {\em \mnras}, 462:3255--3264, November 2016.

\bibitem{aubourg15}
{\'E}.~{Aubourg}, S.~{Bailey}, J.~E. {Bautista}, F.~{Beutler}, V.~{Bhardwaj},
  D.~{Bizyaev}, M.~{Blanton}, M.~{Blomqvist}, A.~S. {Bolton}, J.~{Bovy},
  H.~{Brewington}, J.~{Brinkmann}, J.~R. {Brownstein}, A.~{Burden}, N.~G.
  {Busca}, W.~{Carithers}, C.-H. {Chuang}, J.~{Comparat}, R.~A.~C. {Croft},
  A.~J. {Cuesta}, K.~S. {Dawson}, T.~{Delubac}, D.~J. {Eisenstein},
  A.~{Font-Ribera}, J.~{Ge}, J.-M. {Le Goff}, S.~G.~A. {Gontcho}, J.~R. {Gott},
  J.~E. {Gunn}, H.~{Guo}, J.~{Guy}, J.-C. {Hamilton}, S.~{Ho}, K.~{Honscheid},
  C.~{Howlett}, D.~{Kirkby}, F.~S. {Kitaura}, J.-P. {Kneib}, K.-G. {Lee},
  D.~{Long}, R.~H. {Lupton}, M.~V. {Maga{\~n}a}, V.~{Malanushenko},
  E.~{Malanushenko}, M.~{Manera}, C.~{Maraston}, D.~{Margala}, C.~K. {McBride},
  J.~{Miralda-Escud{\'e}}, A.~D. {Myers}, R.~C. {Nichol}, P.~{Noterdaeme},
  S.~E. {Nuza}, M.~D. {Olmstead}, D.~{Oravetz}, I.~{P{\^a}ris},
  N.~{Padmanabhan}, N.~{Palanque-Delabrouille}, K.~{Pan},
  M.~{Pellejero-Ibanez}, W.~J. {Percival}, P.~{Petitjean}, M.~M. {Pieri},
  F.~{Prada}, B.~{Reid}, J.~{Rich}, N.~A. {Roe}, A.~J. {Ross}, N.~P. {Ross},
  G.~{Rossi}, J.~A. {Rubi{\~n}o-Mart{\'{\i}}n}, A.~G. {S{\'a}nchez},
  L.~{Samushia}, R.~T. {G{\'e}nova-Santos}, C.~G. {Sc{\'o}ccola}, D.~J.
  {Schlegel}, D.~P. {Schneider}, H.-J. {Seo}, E.~{Sheldon}, A.~{Simmons}, R.~A.
  {Skibba}, A.~{Slosar}, M.~A. {Strauss}, D.~{Thomas}, J.~L. {Tinker},
  R.~{Tojeiro}, J.~A. {Vazquez}, M.~{Viel}, D.~A. {Wake}, B.~A. {Weaver}, D.~H.
  {Weinberg}, W.~M. {Wood-Vasey}, C.~{Y{\`e}che}, I.~{Zehavi}, G.-B. {Zhao},
  and {BOSS Collaboration}.
\newblock {Cosmological implications of baryon acoustic oscillation
  measurements}.
\newblock {\em \prd}, 92(12):123516, December 2015.

\bibitem{planck15}
{Planck Collaboration}, P.~A.~R. {Ade}, N.~{Aghanim}, M.~{Arnaud},
  M.~{Ashdown}, J.~{Aumont}, C.~{Baccigalupi}, A.~J. {Banday}, R.~B.
  {Barreiro}, J.~G. {Bartlett}, and et~al.
\newblock {Planck 2015 results. XIII. Cosmological parameters}.
\newblock {\em \aap}, 594:A13, September 2016.

\bibitem{Bonvin17}
V.~{Bonvin}, F.~{Courbin}, S.~H. {Suyu}, P.~J. {Marshall}, C.~E. {Rusu},
  D.~{Sluse}, M.~{Tewes}, K.~C. {Wong}, T.~{Collett}, C.~D. {Fassnacht},
  T.~{Treu}, M.~W. {Auger}, S.~{Hilbert}, L.~V.~E. {Koopmans}, G.~{Meylan},
  N.~{Rumbaugh}, A.~{Sonnenfeld}, and C.~{Spiniello}.
\newblock {H0LiCOW - V. New COSMOGRAIL time delays of HE 0435-1223: H$_{0}$ to
  3.8 per cent precision from strong lensing in a flat {$\Lambda$}CDM model}.
\newblock {\em \mnras}, 465:4914--4930, March 2017.

\bibitem{linder04}
E.~V. {Linder}.
\newblock {Strong gravitational lensing and dark energy complementarity}.
\newblock {\em \prd}, 70(4):043534, August 2004.

\bibitem{linder11}
E.~V. {Linder}.
\newblock {Lensing time delays and cosmological complementarity}.
\newblock {\em \prd}, 84(12):123529, December 2011.

\bibitem{detf_report}
Andreas {Albrecht}, Gary {Bernstein}, Robert {Cahn}, Wendy~L. {Freedman},
  Jacqueline {Hewitt}, Wayne {Hu}, John {Huth}, Marc {Kamionkowski}, Edward~W.
  {Kolb}, Lloyd {Knox}, John~C. {Mather}, Suzanne {Staggs}, and Nicholas~B.
  {Suntzeff}.
\newblock {Report of the Dark Energy Task Force}.
\newblock {\em arXiv e-prints}, pages astro--ph/0609591, Sep 2006.

\bibitem{Snowmass2022:Tensions}
Elcio {Abdalla}, Guillermo~Franco {Abell{\'a}n}, Amin {Aboubrahim}, Adriano
  {Agnello}, Ozgur {Akarsu}, Yashar {Akrami}, George {Alestas}, Daniel {Aloni},
  Luca {Amendola}, Luis~A. {Anchordoqui}, Richard~I. {Anderson}, Nikki
  {Arendse}, Marika {Asgari}, Mario {Ballardini}, Vernon {Barger}, Spyros
  {Basilakos}, Ronaldo~C. {Batista}, Elia~S. {Battistelli}, Richard {Battye},
  Micol {Benetti}, David {Benisty}, Asher {Berlin}, Paolo {de Bernardis},
  Emanuele {Berti}, Bohdan {Bidenko}, Simon {Birrer}, John~P. {Blakeslee},
  Kimberly~K. {Boddy}, Clecio~R. {Bom}, Alexander {Bonilla}, Nicola {Borghi},
  Fran{\c{c}}ois~R. {Bouchet}, Matteo {Braglia}, Thomas {Buchert}, Elizabeth
  {Buckley-Geer}, Erminia {Calabrese}, Robert~R. {Caldwell}, David {Camarena},
  Salvatore {Capozziello}, Stefano {Casertano}, Geoff C.~F. {Chen}, Jens
  {Chluba}, Angela {Chen}, Hsin-Yu {Chen}, Anton {Chudaykin}, Michele {Cicoli},
  Craig~J. {Copi}, Fred {Courbin}, Francis-Yan {Cyr-Racine}, Bozena {Czerny},
  Maria {Dainotti}, Guido {D'Amico}, Anne-Christine {Davis}, Javier {de Cruz
  P{\'e}rez}, Jaume {de Haro}, Jacques {Delabrouille}, Peter~B. {Denton},
  Suhail {Dhawan}, Keith~R. {Dienes}, Eleonora {Di Valentino}, Pu~{Du},
  Dominique {Eckert}, Celia {Escamilla-Rivera}, Agn{\`e}s {Fert{\'e}}, Fabio
  {Finelli}, Pablo {Fosalba}, Wendy~L. {Freedman}, Noemi {Frusciante}, Enrique
  {Gazta{\~n}aga}, William {Giar{\`e}}, Elena {Giusarma}, Adri{\`a}
  {G{\'o}mez-Valent}, Will {Handley}, Ian {Harrison}, Luke {Hart}, Dhiraj~Kumar
  {Hazra}, Alan {Heavens}, Asta {Heinesen}, Hendrik {Hildebrandt}, J.~Colin
  {Hill}, Natalie~B. {Hogg}, Daniel~E. {Holz}, Deanna~C. {Hooper}, Nikoo
  {Hosseininejad}, Dragan {Huterer}, Mustapha {Ishak}, Mikhail~M. {Ivanov},
  Andrew~H. {Jaffe}, In~Sung {Jang}, Karsten {Jedamzik}, Raul {Jimenez},
  Melissa {Joseph}, Shahab {Joudaki}, Mark {Kamionkowski}, Tanvi {Karwal},
  Lavrentios {Kazantzidis}, Ryan~E. {Keeley}, Michael {Klasen}, Eiichiro
  {Komatsu}, L{\'e}on V.~E. {Koopmans}, Suresh {Kumar}, Luca {Lamagna}, Ruth
  {Lazkoz}, Chung-Chi {Lee}, Julien {Lesgourgues}, Jackson {Levi Said},
  Tiffany~R. {Lewis}, Benjamin {L'Huiller}, Matteo {Lucca}, Roy {Maartens},
  Lucas~M. {Macri}, Danny {Marfatia}, Valerio {Marra}, Carlos J.~A.~P.
  {Martins}, Silvia {Masi}, Sabino {Matarrese}, Arindam {Mazumdar}, Alessandro
  {Melchiorri}, Olga {Mena}, Laura {Mersini-Houghton}, James {Mertens}, Dinko
  {Milakovic}, Yuto {Minami}, Vivian {Miranda}, Cristian {Moreno-Pulido},
  Michele {Moresco}, David~F. {Mota}, Emil {Mottola}, Simone {Mozzon}, Jessica
  {Muir}, Ankan {Mukherjee}, Suvodip {Mukherjee}, Pavel {Naselsky}, Pran
  {Nath}, Savvas {Nesseris}, Florian {Niedermann}, Alessio {Notari}, Rafael~C.
  {Nunes}, Eoin~{\'O} {Colg{\'a}in}, Kayla~A. {Owens}, Emre {Ozulker},
  Francesco {Pace}, Andronikos {Paliathanasis}, Antonella {Palmese}, Supriya
  {Pan}, Daniela {Paoletti}, Santiago~E. {Perez Bergliaffa}, Leadros
  {Perivolaropoulos}, Dominic~W. {Pesce}, Valeria {Pettorino}, Oliver H.~E.
  {Philcox}, Levon {Pogosian}, Vivian {Poulin}, Gaspard {Poulot}, Marco
  {Raveri}, Mark~J. {Reid}, Fabrizio {Renzi}, Adam~G. {Riess}, Vivian~I.
  {Sabla}, Paolo {Salucci}, Vincenzo {Salzano}, Emmanuel~N. {Saridakis},
  Bangalore~S. {Sathyaprakash}, Martin {Schmaltz}, Nils {Sch{\"o}neberg}, Dan
  {Scolnic}, Anjan~A. {Sen}, Neelima {Sehgal}, Arman {Shafieloo}, M.~M.
  {Sheikh-Jabbari}, Joseph {Silk}, Alessandra {Silvestri}, Foteini {Skara},
  Martin~S. {Sloth}, Marcelle {Soares-Santos}, Joan {Sol{\`a} Peracaula},
  Yu-Yang {Songsheng}, Jorge~F. {Soriano}, Denitsa {Staicova}, Glenn~D.
  {Starkman}, Istv{\'a}n {Szapudi}, Elsa~M. {Teixera}, Brooks {Thomas}, Tommaso
  {Treu}, Emery {Trott}, Carsten {van de Bruck}, J.~Alberto {Vazquez}, Licia
  {Verde}, Luca {Visinelli}, Deng {Wang}, Jian-Min {Wang}, Shao-Jiang {Wang},
  Richard {Watkins}, Scott {Watson}, John~K. {Webb}, Neal {Weiner}, Amanda
  {Weltman}, Samuel~J. {Witte}, Rados{\l}aw {Wojtak}, Anil~Kumar {Yadav},
  Weiqiang {Yang}, Gong-Bo {Zhao}, and Miguel {Zumalac{\'a}rregui}.
\newblock {Cosmology Intertwined: A Review of the Particle Physics,
  Astrophysics, and Cosmology Associated with the Cosmological Tensions and
  Anomalies}.
\newblock {\em arXiv e-prints}, page arXiv:2203.06142, March 2022.

\bibitem{Astro2020:Beaton}
Rachael~L. {Beaton}, Simon {Birrer}, Ian {Dell'Antonio}, Chris {Fassnacht},
  Danny {Goldstein}, Chien-Hsiu {Lee}, Peter {Nugent}, Michael {Pierce},
  Anowar~J. {Shajib}, and Tommaso {Treu}.
\newblock {Measuring the Hubble Constant Near and Far in the Era of ELT's}.
\newblock {\em \baas}, 51(3):456, May 2019.

\bibitem{2018arXiv180910147G}
D.~A. {Goldstein}, P.~E. {Nugent}, and A.~{Goobar}.
\newblock {Rates and Properties of Strongly Gravitationally Lensed Supernovae
  and their Host Galaxies in Time-Domain Imaging Surveys}.
\newblock {\em arXiv e-prints}, page arXiv:1809.10147, September 2018.

\bibitem{om10}
M.~{Oguri} and P.~J. {Marshall}.
\newblock {Gravitationally lensed quasars and supernovae in future wide-field
  optical imaging surveys}.
\newblock {\em \mnras}, 405:2579--2593, July 2010.

\bibitem{Collett15}
T.~E. {Collett}.
\newblock {The Population of Galaxy-Galaxy Strong Lenses in Forthcoming Optical
  Imaging Surveys}.
\newblock {\em \apj}, 811:20, September 2015.

\bibitem{Goldstein18}
D.~A. {Goldstein}, P.~E. {Nugent}, D.~N. {Kasen}, and T.~E. {Collett}.
\newblock {Precise Time Delays from Strongly Gravitationally Lensed Type Ia
  Supernovae with Chromatically Microlensed Images}.
\newblock {\em \apj}, 855:22, March 2018.

\bibitem{Collett14}
T.~E. {Collett} and M.~W. {Auger}.
\newblock {Cosmological constraints from the double source plane lens
  SDSSJ0946+1006}.
\newblock {\em \mnras}, 443:969--976, September 2014.

\bibitem{Grillo08}
C.~{Grillo}, M.~{Lombardi}, and G.~{Bertin}.
\newblock {Cosmological parameters from strong gravitational lensing and
  stellar dynamics in elliptical galaxies}.
\newblock {\em \aap}, 477:397--406, January 2008.

\bibitem{Birrer18}
S.~{Birrer}, A.~{Refregier}, and A.~{Amara}.
\newblock {Cosmic Shear with Einstein Rings}.
\newblock {\em \apjl}, 852:L14, January 2018.

\bibitem{2014ApJ...788...48S}
B.~J. {Shappee} et~al.
\newblock {The Man behind the Curtain: X-Rays Drive the UV through NIR
  Variability in the 2013 Active Galactic Nucleus Outburst in NGC 2617}.
\newblock {\em \it{ApJ}}.

\bibitem{2018PASP..130f4505T}
J.~L. {Tonry} et~al.
\newblock {ATLAS: A High-cadence All-sky Survey System}.
\newblock {\em \it{Publications of the Astronomical Society of the Pacific}},
  130:064505, Jun 2018.

\bibitem{2020arXiv200109095G}
R.~{Graziani} et~al.
\newblock {Peculiar velocity cosmology with type Ia supernovae}.
\newblock {\em arXiv e-prints}, page arXiv:2001.09095, January 2020.

\bibitem{2018arXiv180901669T}
{The LSST Dark Energy Science Collaboration}.
\newblock {The LSST Dark Energy Science Collaboration (DESC) Science
  Requirements Document}.
\newblock {\em arXiv e-prints}, page arXiv:1809.01669, September 2018.

\bibitem{2018ApJ...859..101S}
D.~M. {Scolnic} et~al.
\newblock {The Complete Light-curve Sample of Spectroscopically Confirmed SNe
  Ia from Pan-STARRS1 and Cosmological Constraints from the Combined Pantheon
  Sample}.
\newblock {\em \it{ApJ}}, 859(2):101, June 2018.

\bibitem{2012MNRAS.425.1007B}
R.~L. {Barone-Nugent} et~al.
\newblock {Near-infrared observations of Type Ia supernovae: the best known
  standard candle for cosmology}.
\newblock {\em \it{MNRAS}}, 425:1007--1012, September 2012.

\bibitem{2018ApJ...869...56B}
Christopher~R. {Burns} et~al.
\newblock {The Carnegie Supernova Project: Absolute Calibration and the Hubble
  Constant}.
\newblock {\em \it{ApJ}}, 869(1):56, December 2018.

\bibitem{2015ApJ...815...58F}
H.~K. {Fakhouri} et~al.
\newblock {Improving Cosmological Distance Measurements Using Twin Type Ia
  Supernovae}.
\newblock {\em ApJ}, 815:58, December 2015.

\bibitem{Boone2020}
K.~Boone et~al.
\newblock {The Twins Embedding of Type Ia Supernovae II: Improving Cosmological
  Distance Estimates}.
\newblock {\em \it{ApJ}}, submitted, 2020.

\bibitem{2019PASP..131k8002M}
Daniel {Muthukrishna}, Gautham {Narayan}, Kaisey~S. {Mandel}, Rahul {Biswas},
  and Ren{\'e}e {Hlo{\v{z}}ek}.
\newblock {RAPID: Early Classification of Explosive Transients Using Deep
  Learning}.
\newblock {\em \it{PASP}}, 131(1005):118002, November 2019.

\bibitem{2018arXiv181200515L}
Michelle {Lochner} et~al.
\newblock {Optimizing the LSST Observing Strategy for Dark Energy Science: DESC
  Recommendations for the Wide-Fast-Deep Survey}.
\newblock {\em arXiv e-prints}, page arXiv:1812.00515, November 2018.

\bibitem{2015A&A...584A..62C}
E.~{Cappellaro} et~al.
\newblock {Supernova rates from the SUDARE VST-OmegaCAM search. I. Rates per
  unit volume}.
\newblock {\em \it{A\&A}}, 584:A62, December 2015.

\bibitem{2004SPIE.5249..146L}
B.~{Lantz} et~al.
\newblock {SNIFS: a wideband integral field spectrograph with microlens
  arrays}.
\newblock {\em Proc. SPIE}, 5249:146--155, February 2004.

\bibitem{Aldering2020}
G.~Aldering et~al.
\newblock in prep., 2022.

\bibitem{2010SPIE.7735E..08B}
R.~{Bacon} et~al.
\newblock {The MUSE second-generation VLT instrument}.
\newblock In {\em Ground-based and Airborne Instrumentation for Astronomy III},
  volume 7735 of {\em Society of Photo-Optical Instrumentation Engineers (SPIE)
  Conference Series}, page 773508, July 2010.

\bibitem{2019BAAS...51g.198B}
Kevin {Bundy} et~al.
\newblock {FOBOS: A Next-Generation Spectroscopic Facility}.
\newblock In {\em Bulletin of the American Astronomical Society}, volume~51,
  page 198, September 2019.

\bibitem{2016arXiv161100036D}
{DESI Collaboration}.
\newblock {The DESI Experiment Part I: Science,Targeting, and Survey Design}.
\newblock {\em arXiv e-prints}, page arXiv:1611.00036, October 2016.

\bibitem{2019eeu..confE..56S}
Elizabeth {Swann}.
\newblock {The Time Domain Extragalactic Survey (TiDES)}.
\newblock In {\em The Extragalactic Explosive Universe: the New Era of
  Transient Surveys and Data-Driven Discovery}, page~56, October 2019.

\bibitem{2018PASP..130c5003B}
Nadejda {Blagorodnova} et~al.
\newblock {The SED Machine: A Robotic Spectrograph for Fast Transient
  Classification}.
\newblock {\em \it{PASP}}, 130(985):035003, March 2018.

\bibitem{2020ApJ...895...32F}
C.~{Fremling} et~al.
\newblock {The Zwicky Transient Facility Bright Transient Survey. I.
  Spectroscopic Classification and the Redshift Completeness of Local Galaxy
  Catalogs}.
\newblock {\em \it{ApJ}}, 895(1):32, May 2020.

\bibitem{hearin_etal10}
A.~P. {Hearin}, A.~R. {Zentner}, Z.~{Ma}, and D.~{Huterer}.
\newblock {A General Study of the Influence of Catastrophic Photometric
  Redshift Errors on Cosmology with Cosmic Shear Tomography}.
\newblock {\em \apj}, 720:1351--1369, September 2010.

\bibitem{detf}
A.~{Albrecht}, G.~{Bernstein}, R.~{Cahn}, W.~L. {Freedman}, J.~{Hewitt},
  W.~{Hu}, J.~{Huth}, M.~{Kamionkowski}, E.~W. {Kolb}, L.~{Knox}, J.~C.
  {Mather}, S.~{Staggs}, and N.~B. {Suntzeff}.
\newblock {Report of the Dark Energy Task Force}.
\newblock {\em arXiv Astrophysics e-prints}, pages arXiv:astro--ph/0609591,
  September 2006.

\bibitem{2006JCAP...08..008Z}
H.~{Zhan}.
\newblock {Cosmic tomographies: baryon acoustic oscillations and weak lensing}.
\newblock {\em Journal of Cosmology and Astro-Particle Physics}, 8:8--+, August
  2006.

\bibitem{2006ApJ...644..663Z}
H.~{Zhan} and L.~{Knox}.
\newblock {Baryon Oscillations and Consistency Tests for Photometrically
  Determined Redshifts of Very Faint Galaxies}.
\newblock {\em \apj}, 644:663--670, June 2006.

\bibitem{2006astro.ph..5536K}
L.~{Knox}, Y.-S. {Song}, and H.~{Zhan}.
\newblock {Weighing the Universe with Photometric Redshift Surveys and the
  Impact on Dark Energy Forecasts}.
\newblock {\em \apj}, 652:857--863, December 2006.

\bibitem{tysonconf}
J.~A. {Tyson}.
\newblock {Precision Studies of Dark Energy with LSST}.
\newblock In T.~M. {Liss}, editor, {\em Intersections of Particle and Nuclear
  Physics: 9th Conference CIPAN2006}, volume 870 of {\em American Institute of
  Physics Conference Series}, pages 44--52, November 2006.

\bibitem{Graham2018}
Melissa~L. {Graham}, Andrew~J. {Connolly}, {\v{Z}}eljko {Ivezi{\'c}}, Samuel~J.
  {Schmidt}, R.~Lynne {Jones}, Mario {Juri{\'c}}, Scott~F. {Daniel}, and Peter
  {Yoachim}.
\newblock {Photometric Redshifts with the LSST: Evaluating Survey Observing
  Strategies}.
\newblock {\em \aj}, 155:1, January 2018.

\bibitem{grahamcat1}
V.~{Springel}, S.~D.~M. {White}, A.~{Jenkins}, C.~S. {Frenk}, N.~{Yoshida},
  L.~{Gao}, J.~{Navarro}, R.~{Thacker}, D.~{Croton}, J.~{Helly}, J.~A.
  {Peacock}, S.~{Cole}, P.~{Thomas}, H.~{Couchman}, A.~{Evrard}, J.~{Colberg},
  and F.~{Pearce}.
\newblock {Simulations of the formation, evolution and clustering of galaxies
  and quasars}.
\newblock {\em \nat}, 435:629--636, June 2005.

\bibitem{grahamcat2}
V.~{Gonzalez-Perez}, C.~G. {Lacey}, C.~M. {Baugh}, C.~D.~P. {Lagos},
  J.~{Helly}, D.~J.~R. {Campbell}, and P.~D. {Mitchell}.
\newblock {How sensitive are predicted galaxy luminosities to the choice of
  stellar population synthesis model?}
\newblock {\em \mnras}, 439:264--283, March 2014.

\bibitem{grahamcat3}
A.~I. {Merson}, C.~M. {Baugh}, J.~C. {Helly}, V.~{Gonzalez-Perez}, S.~{Cole},
  R.~{Bielby}, P.~{Norberg}, C.~S. {Frenk}, A.~J. {Benson}, R.~G. {Bower},
  C.~G. {Lacey}, and C.~d.~P. {Lagos}.
\newblock {Lightcone mock catalogues from semi-analytic models of galaxy
  formation - I. Construction and application to the BzK colour selection}.
\newblock {\em \mnras}, 429:556--578, February 2013.

\bibitem{2015MNRAS.452.3100C}
S.~{Cavuoti}, M.~{Brescia}, C.~{Tortora}, G.~{Longo}, N.~R. {Napolitano},
  M.~{Radovich}, F.~{La Barbera}, M.~{Capaccioli}, J.~T.~A. {de Jong},
  F.~{Getman}, A.~{Grado}, and M.~{Paolillo}.
\newblock {Machine-learning-based photometric redshifts for galaxies of the ESO
  Kilo-Degree Survey data release 2}.
\newblock {\em \mnras}, 452:3100--3105, Sep 2015.

\bibitem{2017ApJ...841..111M}
Daniel~C. {Masters}, Daniel~K. {Stern}, Judith~G. {Cohen}, Peter~L. {Capak},
  Jason~D. {Rhodes}, Francisco~J. {Castander}, and St{\'e}phane {Paltani}.
\newblock {The Complete Calibration of the Color-Redshift Relation (C3R2)
  Survey: Survey Overview and Data Release 1}.
\newblock {\em \apj}, 841:111, Jun 2017.

\bibitem{Samuroff18}
S.~{Samuroff}, J.~{Blazek}, M.~A. {Troxel}, N.~{MacCrann}, E.~{Krause}, C.~D.
  {Leonard}, J.~{Prat}, D.~{Gruen}, S.~{Dodelson}, T.~F. {Eifler}, M.~{Gatti},
  W.~G. {Hartley}, B.~{Hoyle}, P.~{Larsen}, J.~{Zuntz}, T.~M.~C. {Abbott},
  S.~{Allam}, J.~{Annis}, G.~M. {Bernstein}, E.~{Bertin}, S.~L. {Bridle},
  D.~{Brooks}, A.~{Carnero Rosell}, M.~{Carrasco Kind}, J.~{Carretero}, F.~J.
  {Castander}, C.~E. {Cunha}, L.~N. {da Costa}, C.~{Davis}, J.~{De Vicente},
  D.~L. {DePoy}, S.~{Desai}, H.~T. {Diehl}, J.~P. {Dietrich}, P.~{Doel},
  B.~{Flaugher}, P.~{Fosalba}, J.~{Frieman}, J.~{Garc{\'\i}a-Bellido},
  E.~{Gaztanaga}, D.~W. {Gerdes}, R.~A. {Gruendl}, J.~{Gschwend},
  G.~{Gutierrez}, D.~L. {Hollowood}, K.~{Honscheid}, D.~J. {James}, K.~{Kuehn},
  N.~{Kuropatkin}, M.~{Lima}, M.~A.~G. {Maia}, M.~{March}, J.~L. {Marshall},
  P.~{Martini}, P.~{Melchior}, F.~{Menanteau}, C.~J. {Miller}, R.~{Miquel},
  R.~L.~C. {Ogando}, A.~A. {Plazas}, E.~{Sanchez}, V.~{Scarpine},
  R.~{Schindler}, M.~{Schubnell}, S.~{Serrano}, I.~{Sevilla-Noarbe},
  E.~{Sheldon}, M.~{Smith}, F.~{Sobreira}, E.~{Suchyta}, G.~{Tarle},
  D.~{Thomas}, and V.~{Vikram}.
\newblock {Dark Energy Survey Year 1 Results: Constraints on Intrinsic
  Alignments and their Colour Dependence from Galaxy Clustering and Weak
  Lensing}.
\newblock {\em arXiv e-prints}, page arXiv:1811.06989, November 2018.

\bibitem{Sanchez19}
J.~{Sanchez}, I.~{Mendoza}, D.~{Kirkby}, and P.~{Burchat}.
\newblock Olber's paradox revisited -- effects of overlapping sources on cosmic
  shear estimation: Statistical sensitivity and pixel-noise bias, in prep.

\bibitem{Melchior18}
P.~{Melchior}, F.~{Moolekamp}, M.~{Jerdee}, R.~{Armstrong}, A.~L. {Sun},
  J.~{Bosch}, and R.~{Lupton}.
\newblock {SCARLET: Source separation in multi-band images by Constrained
  Matrix Factorization}.
\newblock {\em Astronomy and Computing}, 24:129--142, July 2018.

\bibitem{Euclid19}
{Euclid Collaboration}, N.~{Martinet}, T.~{Schrabback}, H.~{Hoekstra},
  M.~{Tewes}, R.~{Herbonnet}, P.~{Schneider}, B.~{Hernandez-Martin}, A.~N.
  {Taylor}, J.~{Brinchmann}, C.~S. {Carvalho}, M.~{Castellano}, G.~{Congedo},
  B.~R. {Gillis}, E.~{Jullo}, M.~{K{\"u}mmel}, S.~{Ligori}, P.~B. {Lilje},
  C.~{Padilla}, D.~{Paris}, J.~A. {Peacock}, S.~{Pilo}, A.~{Pujol}, D.~{Scott},
  and R.~{Toledo-Moreo}.
\newblock {Euclid Preparation IV. Impact of undetected galaxies on weak lensing
  shear measurements}.
\newblock {\em arXiv e-prints}, page arXiv:1902.00044, January 2019.

\bibitem{Gruen18}
D.~{Gruen}, Y.~{Zhang}, A.~{Palmese}, B.~{Yanny}, V.~{Busti}, B.~{Hoyle},
  P.~{Melchior}, C.~J. {Miller}, E.~{Rozo}, E.~S. {Rykoff}, T.~N. {Varga},
  F.~B. {Abdalla}, S.~{Allam}, J.~{Annis}, S.~{Avila}, D.~{Brooks}, D.~L.
  {Burke}, A.~{Carnero Rosell}, M.~{Carrasco Kind}, J.~{Carretero},
  R.~{Cawthon}, M.~{Crocce}, C.~E. {Cunha}, L.~N. {da Costa}, C.~{Davis},
  J.~{De Vicente}, S.~{Desai}, H.~T. {Diehl}, J.~P. {Dietrich},
  A.~{Drlica-Wagner}, B.~{Flaugher}, P.~{Fosalba}, J.~{Frieman},
  J.~{Garc{\'\i}a-Bellido}, E.~{Gaztanaga}, D.~W. {Gerdes}, R.~A. {Gruendl},
  J.~{Gschwend}, D.~L. {Hollowood}, K.~{Honscheid}, D.~J. {James},
  T.~{Jeltema}, E.~{Krause}, R.~{Kron}, K.~{Kuehn}, N.~{Kuropatkin},
  O.~{Lahav}, M.~{Lima}, H.~{Lin}, M.~A.~G. {Maia}, J.~L. {Marshall},
  F.~{Menanteau}, R.~{Miquel}, R.~L.~C. {Ogando}, A.~A. {Plazas}, A.~K.
  {Romer}, V.~{Scarpine}, I.~{Sevilla-Noarbe}, M.~{Smith}, M.~{Soares-Santos},
  F.~{Sobreira}, E.~{Suchyta}, M.~E.~C. {Swanson}, G.~{Tarle}, D.~{Thomas},
  V.~{Vikram}, and A.~R. {Walker}.
\newblock {Dark Energy Survey Year 1 Results: The effect of intra-cluster light
  on photometric redshifts for weak gravitational lensing}.
\newblock {\em arXiv e-prints}, page arXiv:1809.04599, September 2018.

\bibitem{Newman2013}
Jeffrey~A. {Newman}, Michael~C. {Cooper}, Marc {Davis}, S.~M. {Faber},
  Alison~L. {Coil}, Puragra {Guhathakurta}, David~C. {Koo}, Andrew~C.
  {Phillips}, Charlie {Conroy}, Aaron~A. {Dutton}, Douglas~P. {Finkbeiner},
  Brian~F. {Gerke}, David~J. {Rosario}, Benjamin~J. {Weiner}, C.~N.~A.
  {Willmer}, Renbin {Yan}, Justin~J. {Harker}, Susan~A. {Kassin}, N.~P.
  {Konidaris}, Kamson {Lai}, Darren~S. {Madgwick}, K.~G. {Noeske}, Gregory~D.
  {Wirth}, A.~J. {Connolly}, N.~{Kaiser}, Evan~N. {Kirby}, Brian~C. {Lemaux},
  Lihwai {Lin}, Jennifer~M. {Lotz}, G.~A. {Luppino}, C.~{Marinoni}, Daniel~J.
  {Matthews}, Anne {Metevier}, and Ricardo~P. {Schiavon}.
\newblock {The DEEP2 Galaxy Redshift Survey: Design, Observations, Data
  Reduction, and Redshifts}.
\newblock {\em \apjs}, 208(1):5, September 2013.

\bibitem{Lilly2007}
S.~J. {Lilly}, O.~{Le F{\`e}vre}, A.~{Renzini}, G.~{Zamorani}, M.~{Scodeggio},
  T.~{Contini}, C.~M. {Carollo}, G.~{Hasinger}, J.~P. {Kneib}, A.~{Iovino},
  V.~{Le Brun}, C.~{Maier}, V.~{Mainieri}, M.~{Mignoli}, J.~{Silverman},
  L.~A.~M. {Tasca}, M.~{Bolzonella}, A.~{Bongiorno}, D.~{Bottini}, P.~{Capak},
  K.~{Caputi}, A.~{Cimatti}, O.~{Cucciati}, E.~{Daddi}, R.~{Feldmann},
  P.~{Franzetti}, B.~{Garilli}, L.~{Guzzo}, O.~{Ilbert}, and P.~{Kampczyk}.
\newblock {zCOSMOS: A Large VLT/VIMOS Redshift Survey Covering 0 < z < 3 in the
  COSMOS Field}.
\newblock {\em \apjs}, 172(1):70--85, September 2007.

\bibitem{2008ApJ...684...88N}
J.~A. {Newman}.
\newblock {Calibrating Redshift Distributions beyond Spectroscopic Limits with
  Cross-Correlations}.
\newblock {\em \apj}, 684:88--101, September 2008.

\bibitem{Schmidt:2013sba}
Samuel~J. {Schmidt}, Brice {M{\'e}nard}, Ryan {Scranton}, Christopher
  {Morrison}, and Cameron~K. {McBride}.
\newblock {Recovering redshift distributions with cross-correlations: pushing
  the boundaries}.
\newblock {\em \mnras}, 431:3307--3318, Jun 2013.

\bibitem{Rahman:2014lfa}
Mubdi {Rahman}, Brice {M{\'e}nard}, Ryan {Scranton}, Samuel~J. {Schmidt}, and
  Christopher~B. {Morrison}.
\newblock {Clustering-based redshift estimation: comparison to spectroscopic
  redshifts}.
\newblock {\em \mnras}, 447:3500--3511, Mar 2015.

\bibitem{4most}
R.~S. {de Jong}, O.~{Agertz}, A.~Agudo {Berbel}, J.~{Aird}, D.~A. {Alexander},
  A.~{Amarsi}, F.~{Anders}, R.~{Andrae}, B.~{Ansarinejad}, W.~{Ansorge},
  P.~{Antilogus}, H.~{Anwand -Heerwart}, A.~{Arentsen}, A.~{Arnadottir},
  M.~{Asplund}, M.~{Auger}, N.~{Azais}, D.~{Baade}, G.~{Baker}, S.~{Baker},
  E.~{Balbinot}, I.~K. {Baldry}, M.~{Banerji}, S.~{Barden}, P.~{Barklem},
  E.~{Barth{\'e}l{\'e}my-Mazot}, C.~{Battistini}, S.~{Bauer}, C.~P.~M. {Bell},
  O.~{Bellido-Tirado}, S.~{Bellstedt}, V.~{Belokurov}, T.~{Bensby},
  M.~{Bergemann}, J.~M. {Bestenlehner}, R.~{Bielby}, M.~{Bilicki}, C.~{Blake},
  J.~{Bland-Hawthorn}, C.~{Boeche}, W.~{Boland}, T.~{Boller}, S.~{Bongard},
  A.~{Bongiorno}, P.~{Bonifacio}, D.~{Boudon}, D.~{Brooks}, M.~J.~I. {Brown},
  R.~{Brown}, M.~{Br{\"u}ggen}, J.~{Brynnel}, J.~{Brzeski}, T.~{Buchert},
  P.~{Buschkamp}, E.~{Caffau}, P.~{Caillier}, J.~{Carrick}, L.~{Casagrande},
  S.~{Case}, A.~{Casey}, I.~{Cesarini}, G.~{Cescutti}, D.~{Chapuis},
  C.~{Chiappini}, M.~{Childress}, N.~{Christlieb}, R.~{Church}, M.~R.~L.
  {Cioni}, M.~{Cluver}, M.~{Colless}, T.~{Collett}, J.~{Comparat}, A.~{Cooper},
  W.~{Couch}, F.~{Courbin}, S.~{Croom}, D.~{Croton}, E.~{Daguis{\'e}},
  G.~{Dalton}, L.~J.~M. {Davies}, T.~{Davis}, P.~{de Laverny}, A.~{Deason},
  F.~{Dionies}, K.~{Disseau}, P.~{Doel}, D.~{D{\"o}scher}, S.~P. {Driver},
  T.~{Dwelly}, D.~{Eckert}, A.~{Edge}, B.~{Edvardsson}, D.~El {Youssoufi},
  A.~{Elhaddad}, H.~{Enke}, G.~{Erfanianfar}, T.~{Farrell}, T.~{Fechner},
  C.~{Feiz}, S.~{Feltzing}, I.~{Ferreras}, D.~{Feuerstein}, D.~{Feuillet},
  A.~{Finoguenov}, D.~{Ford}, S.~{Fotopoulou}, M.~{Fouesneau}, C.~{Frenk},
  S.~{Frey}, W.~{Gaessler}, S.~{Geier}, N.~Gentile {Fusillo}, O.~{Gerhard},
  T.~{Giannantonio}, D.~{Giannone}, B.~{Gibson}, P.~{Gillingham},
  C.~{Gonz{\'a}lez-Fern{\'a}ndez}, E.~{Gonzalez-Solares}, S.~{Gottloeber},
  A.~{Gould}, E.~K. {Grebel}, A.~{Gueguen}, G.~{Guiglion}, M.~{Haehnelt},
  T.~{Hahn}, C.~J. {Hansen}, H.~{Hartman}, K.~{Hauptner}, K.~{Hawkins},
  D.~{Haynes}, R.~{Haynes}, U.~{Heiter}, A.~{Helmi}, C.~Hernandez {Aguayo},
  P.~{Hewett}, S.~{Hinton}, D.~{Hobbs}, S.~{Hoenig}, D.~{Hofman}, I.~{Hook},
  J.~{Hopgood}, A.~{Hopkins}, A.~{Hourihane}, L.~{Howes}, C.~{Howlett},
  T.~{Huet}, M.~{Irwin}, O.~{Iwert}, P.~{Jablonka}, T.~{Jahn}, K.~{Jahnke},
  A.~{Jarno}, S.~{Jin}, P.~{Jofre}, D.~{Johl}, D.~{Jones}, H.~{J{\"o}nsson},
  C.~{Jordan}, I.~{Karovicova}, A.~{Khalatyan}, A.~{Kelz}, R.~{Kennicutt},
  D.~{King}, F.~{Kitaura}, J.~{Klar}, U.~{Klauser}, J.~{Kneib}, A.~{Koch},
  S.~{Koposov}, G.~{Kordopatis}, A.~{Korn}, J.~{Kosmalski}, R.~{Kotak},
  M.~{Kovalev}, K.~{Kreckel}, Y.~{Kripak}, M.~{Krumpe}, K.~{Kuijken},
  A.~{Kunder}, I.~{Kushniruk}, M.~I {Lam}, G.~{Lamer}, F.~{Laurent},
  J.~{Lawrence}, M.~{Lehmitz}, B.~{Lemasle}, J.~{Lewis}, B.~{Li}, C.~{Lidman},
  K.~{Lind}, J.~{Liske}, J.~L. {Lizon}, J.~{Loveday}, H.~G. {Ludwig}, R.~M.
  {McDermid}, K.~{Maguire}, V.~{Mainieri}, S.~{Mali}, H.~{Mandel}, K.~{Mandel},
  L.~{Mannering}, S.~{Martell}, D.~Martinez {Delgado}, G.~{Matijevic},
  H.~{McGregor}, R.~{McMahon}, P.~{McMillan}, O.~{Mena}, A.~{Merloni}, M.~J.
  {Meyer}, C.~{Michel}, G.~{Micheva}, J.~E. {Migniau}, I.~{Minchev},
  G.~{Monari}, R.~{Muller}, D.~{Murphy}, D.~{Muthukrishna}, K.~{Nandra},
  R.~{Navarro}, M.~{Ness}, V.~{Nichani}, R.~{Nichol}, H.~{Nicklas},
  F.~{Niederhofer}, P.~{Norberg}, D.~{Obreschkow}, S.~{Oliver}, M.~{Owers},
  N.~{Pai}, S.~{Pankratow}, D.~{Parkinson}, I.~{Parry}, J.~{Paschke},
  R.~{Paterson}, A.~{Pecontal}, D.~{Phillips}, A.~{Pillepich}, L.~{Pinard},
  J.~{Pirard}, N.~{Piskunov}, V.~{Plank}, D.~{Pl{\"u}schke}, E.~{Pons},
  P.~{Popesso}, C.~{Power}, J.~{Pragt}, A.~{Pramskiy}, D.~{Pryer},
  M.~{Quattri}, A.~B. de~Andrade {Queiroz}, A.~{Quirrenbach}, S.~{Rahurkar},
  A.~{Raichoor}, S.~{Ramstedt}, A.~{Rau}, A.~{Recio-Blanco}, R.~{Reiss},
  F.~{Renaud}, Y.~{Revaz}, P.~{Rhode}, J.~{Richard}, A.~D. {Richter}, H.~W.
  {Rix}, A.~S.~G. {Robotham}, R.~{Roelfsema}, M.~{Romaniello}, D.~{Rosario},
  F.~{Rothmaier}, B.~{Roukema}, G.~{Ruchti}, G.~{Rupprecht}, J.~{Rybizki},
  N.~{Ryde}, A.~{Saar}, E.~{Sadler}, M.~{Sahl{\'e}n}, M.~{Salvato},
  B.~{Sassolas}, W.~{Saunders}, A.~{Saviauk}, L.~{Sbordone}, T.~{Schmidt},
  O.~{Schnurr}, R.~D. {Scholz}, A.~{Schwope}, W.~{Seifert}, T.~{Shanks},
  A.~{Sheinis}, T.~{Sivov}, {\'A}.~{Sk{\'u}lad{\'o}ttir}, S.~{Smartt},
  S.~{Smedley}, G.~{Smith}, R.~{Smith}, J.~{Sorce}, L.~{Spitler},
  E.~{Starkenburg}, M.~{Steinmetz}, I.~{Stilz}, J.~{Storm}, M.~{Sullivan},
  W.~{Sutherland}, E.~{Swann}, A.~{Tamone}, E.~N. {Taylor}, J.~{Teillon},
  E.~{Tempel}, R.~{ter Horst}, W.~F. {Thi}, E.~{Tolstoy}, S.~{Trager},
  G.~{Traven}, P.~E. {Tremblay}, L.~{Tresse}, M.~{Valentini}, R.~{van de
  Weygaert}, M.~van~den {Ancker}, J.~{Veljanoski}, S.~{Venkatesan},
  L.~{Wagner}, K.~{Wagner}, C.~J. {Walcher}, L.~{Waller}, N.~{Walton},
  L.~{Wang}, R.~{Winkler}, L.~{Wisotzki}, C.~C. {Worley}, G.~{Worseck},
  M.~{Xiang}, W.~{Xu}, D.~{Yong}, C.~{Zhao}, J.~{Zheng}, F.~{Zscheyge}, and
  D.~{Zucker}.
\newblock {4MOST: Project overview and information for the First Call for
  Proposals}.
\newblock {\em arXiv e-prints}, page arXiv:1903.02464, Mar 2019.

\bibitem{joachimi15rev}
B.~{Joachimi}, M.~{Cacciato}, T.~D. {Kitching}, A.~{Leonard}, R.~{Mandelbaum},
  B.~M. {Sch{\"a}fer}, C.~{Sif{\'o}n}, H.~{Hoekstra}, A.~{Kiessling},
  D.~{Kirk}, and A.~{Rassat}.
\newblock {Galaxy Alignments: An Overview}.
\newblock {\em \ssr}, 193:1--65, November 2015.

\bibitem{Troxel:2014dba}
M.~A. Troxel and Mustapha Ishak.
\newblock {The Intrinsic Alignment of Galaxies and its Impact on Weak
  Gravitational Lensing in an Era of Precision Cosmology}.
\newblock {\em Phys. Rept.}, 558:1--59, 2014.

\bibitem{Kirk12}
D.~{Kirk}, A.~{Rassat}, O.~{Host}, and S.~{Bridle}.
\newblock {The cosmological impact of intrinsic alignment model choice for
  cosmic shear}.
\newblock {\em \mnras}, 424:1647--1657, August 2012.

\bibitem{Krause16}
Elisabeth {Krause}, Tim {Eifler}, and Jonathan {Blazek}.
\newblock {The impact of intrinsic alignment on current and future cosmic shear
  surveys}.
\newblock {\em \mnras}, 456:207--222, February 2016.

\bibitem{blazek19}
Jonathan~A. Blazek, Niall MacCrann, M.~A. Troxel, and Xiao Fang.
\newblock Beyond linear galaxy alignments.
\newblock {\em Phys. Rev. D}, 100:103506, Nov 2019.

\bibitem{Yao17}
J.~{Yao}, M.~{Ishak}, W.~{Lin}, and M.~{Troxel}.
\newblock {Effects of self-calibration of intrinsic alignment on cosmological
  parameter constraints from future cosmic shear surveys}.
\newblock {\em \jcap}, 10:056, October 2017.

\bibitem{martens18}
Daniel {Martens}, Christopher~M. {Hirata}, Ashley~J. {Ross}, and Xiao {Fang}.
\newblock {A radial measurement of the galaxy tidal alignment magnitude with
  BOSS data}.
\newblock {\em \mnras}, 478(1):711--732, July 2018.

\bibitem{Chisari16}
Nora~Elisa {Chisari}, Cora {Dvorkin}, Fabian {Schmidt}, and David~N. {Spergel}.
\newblock {Multitracing anisotropic non-Gaussianity with galaxy shapes}.
\newblock {\em \prd}, 94:123507, December 2016.

\bibitem{Joachimi11}
B.~{Joachimi}, R.~{Mandelbaum}, F.~B. {Abdalla}, and S.~L. {Bridle}.
\newblock {Constraints on intrinsic alignment contamination of weak lensing
  surveys using the MegaZ-LRG sampl\ e}.
\newblock {\em \aap}, 527:A26, March 2011.

\bibitem{Johnston18}
Harry {Johnston}, Christos {Georgiou}, Benjamin {Joachimi}, Henk {Hoekstra},
  Nora~Elisa {Chisari}, Daniel {Farrow}, Maria~Cristina {Fortuna}, Catherine
  {Heymans}, Shahab {Joudaki}, Konrad {Kuijken}, and Angus {Wright}.
\newblock {KiDS+GAMA: Intrinsic alignment model constraints for current and
  future weak lensing cosmology}.
\newblock {\em arXiv e-prints}, page arXiv:1811.09598, November 2018.

\bibitem{Tenneti15}
A.~{Tenneti}, R.~{Mandelbaum}, and T.~{Di Matteo}.
\newblock {Intrinsic alignments of disk and elliptical galaxies in the
  MassiveBlack-II and Illustris simulati\ ons}.
\newblock {\em ArXiv e-prints}, October 2015.

\bibitem{Chisari15}
N.~{Chisari}, S.~{Codis}, C.~{Laigle}, Y.~{Dubois}, C.~{Pichon}, and
  J.~{Devriendt}.
\newblock {Intrinsic alignments of galaxies in the Horizon-AGN cosmological
  hydrodynamical simulation}.
\newblock {\em \mnras}, 454:2736--2753, December 2015.

\bibitem{fortuna21a}
Maria~Cristina {Fortuna}, Henk {Hoekstra}, Benjamin {Joachimi}, Harry
  {Johnston}, Nora~Elisa {Chisari}, Christos {Georgiou}, and Constance
  {Mahony}.
\newblock {The halo model as a versatile tool to predict intrinsic alignments}.
\newblock {\em \mnras}, 501(2):2983--3002, February 2021.

\bibitem{Johnston19}
Harry {Johnston}, Christos {Georgiou}, Benjamin {Joachimi}, Henk {Hoekstra},
  Nora~Elisa {Chisari}, Daniel {Farrow}, Maria~Cristina {Fortuna}, Catherine
  {Heymans}, Shahab {Joudaki}, Konrad {Kuijken}, and Angus {Wright}.
\newblock {KiDS+GAMA: Intrinsic alignment model constraints for current and
  future weak lensing cosmology}.
\newblock {\em \aap}, 624:A30, Apr 2019.

\bibitem{Singh14}
S.~{Singh}, R.~{Mandelbaum}, and S.~{More}.
\newblock {Intrinsic alignments of SDSS-III BOSS LOWZ sample galaxies}.
\newblock {\em \mnras}, 450:2195--2216, June 2015.

\bibitem{2014MNRAS.445..726C}
N.~E. {Chisari}, R.~{Mandelbaum}, M.~A. {Strauss}, E.~M. {Huff}, and N.~A.
  {Bahcall}.
\newblock {Intrinsic alignments of group and cluster galaxies in photometric
  surveys}.
\newblock {\em \mnras}, 445:726--748, November 2014.

\bibitem{Kraljic18}
K.~{Kraljic}, S.~{Arnouts}, C.~{Pichon}, C.~{Laigle}, S.~{de la Torre},
  D.~{Vibert}, C.~{Cadiou}, Y.~{Dubois}, M.~{Treyer}, C.~{Schimd}, S.~{Codis},
  V.~{de Lapparent}, J.~{Devriendt}, H.~S. {Hwang}, D.~{Le Borgne},
  N.~{Malavasi}, B.~{Milliard}, M.~{Musso}, D.~{Pogosyan}, M.~{Alpaslan},
  J.~{Bland-Hawthorn}, and A.~H. {Wright}.
\newblock {Galaxy evolution in the metric of the cosmic web}.
\newblock {\em \mnras}, 474:547--571, February 2018.

\bibitem{Codis18}
S.~{Codis}, A.~{Jindal}, N.~E. {Chisari}, D.~{Vibert}, Y.~{Dubois},
  C.~{Pichon}, and J.~{Devriendt}.
\newblock {Galaxy orientation with the cosmic web across cosmic time}.
\newblock {\em \mnras}, 481:4753--4774, December 2018.

\bibitem{BigSky}
Knut {Olsen}, Marcella {Di Criscienzo}, R.~Lynne {Jones}, Megan~E. {Schwamb},
  Hsing~Wen {''Edward'' Lin}, Humna {Awan}, Phil {Marshall}, Eric {Gawiser},
  Adam {Bolton}, and Daniel {Eisenstein}.
\newblock {A Big Sky Approach to Cadence Diplomacy}.
\newblock {\em arXiv e-prints}, page arXiv:1812.02204, Dec 2018.

\bibitem{Schmidt15}
F.~{Schmidt}, N.~E. {Chisari}, and C.~{Dvorkin}.
\newblock {Imprint of inflation on galaxy shape correlations}.
\newblock {\em ArXiv e-prints}, June 2015.

\bibitem{WtG3}
D.~E. {Applegate}, A.~{von der Linden}, P.~L. {Kelly}, M.~T. {Allen}, S.~W.
  {Allen}, P.~R. {Burchat}, D.~L. {Burke}, H.~{Ebeling}, A.~{Mantz}, and R.~G.
  {Morris}.
\newblock {Weighing the Giants - III. Methods and measurements of accurate
  galaxy cluster weak-lensing masses}.
\newblock {\em \mnras}, 439:48--72, March 2014.

\bibitem{McClintock19}
T.~{McClintock}, T.~N. {Varga}, D.~{Gruen}, E.~{Rozo}, E.~S. {Rykoff},
  T.~{Shin}, P.~{Melchior}, J.~{DeRose}, S.~{Seitz}, J.~P. {Dietrich},
  E.~{Sheldon}, Y.~{Zhang}, A.~{von der Linden}, T.~{Jeltema}, A.~B. {Mantz},
  A.~K. {Romer}, S.~{Allen}, M.~R. {Becker}, A.~{Bermeo}, S.~{Bhargava},
  M.~{Costanzi}, S.~{Everett}, A.~{Farahi}, N.~{Hamaus}, W.~G. {Hartley}, D.~L.
  {Hollowood}, B.~{Hoyle}, H.~{Israel}, P.~{Li}, N.~{MacCrann}, G.~{Morris},
  A.~{Palmese}, A.~A. {Plazas}, G.~{Pollina}, M.~M. {Rau}, M.~{Simet},
  M.~{Soares-Santos}, M.~A. {Troxel}, C.~{Vergara Cervantes}, R.~H. {Wechsler},
  J.~{Zuntz}, T.~M.~C. {Abbott}, F.~B. {Abdalla}, S.~{Allam}, J.~{Annis},
  S.~{Avila}, S.~L. {Bridle}, D.~{Brooks}, D.~L. {Burke}, A.~{Carnero Rosell},
  M.~{Carrasco Kind}, J.~{Carretero}, F.~J. {Castander}, M.~{Crocce}, C.~E.
  {Cunha}, C.~B. {D'Andrea}, L.~N. {da Costa}, C.~{Davis}, J.~{De Vicente},
  H.~T. {Diehl}, P.~{Doel}, A.~{Drlica-Wagner}, A.~E. {Evrard}, B.~{Flaugher},
  P.~{Fosalba}, J.~{Frieman}, J.~{Garc{\'{\i}}a-Bellido}, E.~{Gaztanaga}, D.~W.
  {Gerdes}, T.~{Giannantonio}, R.~A. {Gruendl}, G.~{Gutierrez}, K.~{Honscheid},
  D.~J. {James}, D.~{Kirk}, E.~{Krause}, K.~{Kuehn}, O.~{Lahav}, T.~S. {Li},
  M.~{Lima}, M.~{March}, J.~L. {Marshall}, F.~{Menanteau}, R.~{Miquel}, J.~J.
  {Mohr}, B.~{Nord}, R.~L.~C. {Ogando}, A.~{Roodman}, E.~{Sanchez},
  V.~{Scarpine}, R.~{Schindler}, I.~{Sevilla-Noarbe}, M.~{Smith}, R.~C.
  {Smith}, F.~{Sobreira}, E.~{Suchyta}, M.~E.~C. {Swanson}, G.~{Tarle}, D.~L.
  {Tucker}, V.~{Vikram}, A.~R. {Walker}, and J.~{Weller}.
\newblock {Dark Energy Survey Year 1 results: weak lensing mass calibration of
  redMaPPer galaxy clusters}.
\newblock {\em \mnras}, 482:1352--1378, January 2019.

\bibitem{CLASH_photoz}
A.~{Molino}, N.~{Ben{\'{\i}}tez}, B.~{Ascaso}, D.~{Coe}, M.~{Postman},
  S.~{Jouvel}, O.~{Host}, O.~{Lahav}, S.~{Seitz}, E.~{Medezinski}, P.~{Rosati},
  W.~{Schoenell}, A.~{Koekemoer}, Y.~{Jimenez-Teja}, T.~{Broadhurst},
  P.~{Melchior}, I.~{Balestra}, M.~{Bartelmann}, R.~{Bouwens}, L.~{Bradley},
  N.~{Czakon}, M.~{Donahue}, H.~{Ford}, O.~{Graur}, G.~{Graves}, C.~{Grillo},
  L.~{Infante}, S.~W. {Jha}, D.~{Kelson}, R.~{Lazkoz}, D.~{Lemze}, D.~{Maoz},
  A.~{Mercurio}, M.~{Meneghetti}, J.~{Merten}, L.~{Moustakas}, M.~{Nonino},
  S.~{Orgaz}, A.~{Riess}, S.~{Rodney}, J.~{Sayers}, K.~{Umetsu}, W.~{Zheng},
  and A.~{Zitrin}.
\newblock {CLASH: accurate photometric redshifts with 14 HST bands in massive
  galaxy cluster cores}.
\newblock {\em \mnras}, 470:95--113, September 2017.

\bibitem{Diaferio99}
A.~{Diaferio}.
\newblock {Mass estimation in the outer regions of galaxy clusters}.
\newblock {\em \mnras}, 309:610--622, November 1999.

\bibitem{Rines03}
K.~{Rines}, M.~J. {Geller}, M.~J. {Kurtz}, and A.~{Diaferio}.
\newblock {CAIRNS: The Cluster and Infall Region Nearby Survey. I. Redshifts
  and Mass Profiles}.
\newblock {\em \aj}, 126:2152--2170, November 2003.

\bibitem{Falco13}
M.~{Falco}, G.~A. {Mamon}, R.~{Wojtak}, S.~H. {Hansen}, and S.~{Gottl{\"o}ber}.
\newblock {Dynamical signatures of infall around galaxy clusters: a generalized
  Jeans equation}.
\newblock {\em \mnras}, 436:2639--2649, December 2013.

\bibitem{Arthur17}
J.~{Arthur}, F.~R. {Pearce}, M.~E. {Gray}, P.~J. {Elahi}, A.~{Knebe}, A.~M.
  {Beck}, W.~{Cui}, D.~{Cunnama}, R.~{Dav{\'e}}, S.~{February}, S.~{Huang},
  N.~{Katz}, S.~T. {Kay}, I.~G. {McCarthy}, G.~{Murante}, V.~{Perret},
  C.~{Power}, E.~{Puchwein}, A.~{Saro}, F.~{Sembolini}, R.~{Teyssier}, and
  G.~{Yepes}.
\newblock {nIFTy galaxy cluster simulations - V. Investigation of the cluster
  infall region}.
\newblock {\em \mnras}, 464:2027--2038, January 2017.

\bibitem{Rines18}
K.~J. {Rines}, M.~J. {Geller}, A.~{Diaferio}, H.~S. {Hwang}, and J.~{Sohn}.
\newblock {HeCS-red: Dense Hectospec Surveys of redMaPPer-selected Clusters}.
\newblock {\em \apj}, 862:172, August 2018.

\bibitem{Zu14}
Y.~{Zu}, D.~H. {Weinberg}, E.~{Jennings}, B.~{Li}, and M.~{Wyman}.
\newblock {Galaxy infall kinematics as a test of modified gravity}.
\newblock {\em \mnras}, 445:1885--1897, December 2014.

\bibitem{CMB-S4:2016ple}
Kevork~N. Abazajian et~al.
\newblock {CMB-S4 Science Book, First Edition}.
\newblock 10 2016.

\bibitem{Abazajian:2019eic}
{K. Abazajian \textit{et al.} (CMB-S4 Collaboration)}.
\newblock {CMB-S4 Science Case, Reference Design and Project Plan}.
\newblock 2019.

\bibitem{Sunyaev:1970eu}
R.~A. Sunyaev and Ya.~B. Zeldovich.
\newblock {Small scale fluctuations of relic radiation}.
\newblock {\em Astrophys. Space Sci.}, 7:3--19, 1970.

\bibitem{Carlstrom:2002}
J.~E. {Carlstrom}, G.~P. {Holder}, and E.~D. {Reese}.
\newblock {Cosmology with the Sunyaev-Zel'dovich Effect}.
\newblock {\em \araa}, 40:643--680, 2002.

\bibitem{Baldauf2010}
T.~{Baldauf}, R.~E. {Smith}, U.~{Seljak}, and R.~{Mandelbaum}.
\newblock {Algorithm for the direct reconstruction of the dark matter
  correlation function from weak lensing and galaxy clustering}.
\newblock {\em \prd}, 81(6):063531, March 2010.

\bibitem{Mandelbaum2013}
R.~{Mandelbaum}, A.~{Slosar}, T.~{Baldauf}, U.~{Seljak}, C.~M. {Hirata},
  R.~{Nakajima}, R.~{Reyes}, and R.~E. {Smith}.
\newblock {Cosmological parameter constraints from galaxy-galaxy lensing and
  galaxy clustering with the SDSS DR7}.
\newblock {\em \mnras}, 432:1544--1575, June 2013.

\bibitem{Reyes2010}
R.~{Reyes}, R.~{Mandelbaum}, U.~{Seljak}, T.~{Baldauf}, J.~E. {Gunn},
  L.~{Lombriser}, and R.~E. {Smith}.
\newblock {Confirmation of general relativity on large scales from weak lensing
  and galaxy velocities}.
\newblock {\em \nat}, 464:256--258, March 2010.

\bibitem{Zhang2007}
P.~{Zhang}, M.~{Liguori}, R.~{Bean}, and S.~{Dodelson}.
\newblock {Probing Gravity at Cosmological Scales by Measurements which Test
  the Relationship between Gravitational Lensing and Matter Overdensity}.
\newblock {\em Physical Review Letters}, 99(14):141302, October 2007.

\bibitem{Capak2019}
P.~{Capak}, D.~{Sconlic}, J-C. {Cuillandre}, F.~{Castander}, A.~{Bolton},
  R.~{Bowler}, C.~{Chang}, A.~{Dey}, T.~{Eifler}, D.~{Eisenstein},
  C.~{Grillmair}, P.~{Gris}, N.~{Hernitschek}, I.~{Hook}, C.~{Hirata},
  B.~Jain~K. {Kuijken}, M.~{Lochner}, J.~{Newman}, P.~{Oesch}, K.~{Olsen},
  J.~{Rhodes}, B.~{Robertson}, D.~{Rubin}, C.~{Scarlata}, J.~{Silverman},
  S.~{Wachter}, Y.~{Wang}, and {The Tri-Agency Working Group}.
\newblock {Mini-survey of the northern sky to Dec <+30}.
\newblock {\em arXiv e-prints}, page arXiv:1904.10438, April 2019.

\bibitem{megamapper}
David {Schlegel}, Juna~A. {Kollmeier}, and Simone {Ferraro}.
\newblock {The MegaMapper: a z>2 spectroscopic instrument for the study of
  Inflation and Dark Energy}.
\newblock In {\em Bulletin of the American Astronomical Society}, volume~51,
  page 229, September 2019.

\bibitem{mse}
Alexis {Hill}, Nicolas {Flagey}, Alan {McConnachie}, Kei {Szeto}, Andre
  {Anthony}, Javier {Ari{\~n}o}, Ferdinand {Babas}, Gregoire {Bagnoud},
  Gabriella {Baker}, Gregory {Barrick}, Steve {Bauman}, Tom {Benedict},
  Christophe {Berthod}, Armando {Bilbao}, Alberto {Bizkarguenaga}, Alexandre
  {Blin}, Colin {Bradley}, Denis {Brousseau}, Rebecca {Brown}, Jurek {Brzeski},
  Walter {Brzezik}, Patrick {Caillier}, Ram{\'o}n {Campo}, Pierre-Henri
  {Carton}, Jiaru {Chu}, Vladimir {Churilov}, David {Crampton}, Lisa {Crofoot},
  Laurie {Dale}, Lander {de Bilbao}, Markel {Sainz de la Maza}, Daniel
  {Devost}, Michael {Edgar}, Darren {Erickson}, Tony {Farrell}, Pascal
  {Fouque}, Paul {Fournier}, Javier {Garrido}, Mike {Gedig}, Nicolas
  {Geyskens}, James {Gilbert}, Peter {Gillingham}, Guillermo {Gonz{\'a}lez de
  Rivera}, Greg {Green}, Eric {Grigel}, Patrick {Hall}, Kevin {Ho}, David
  {Horville}, Hongzhuan {Hu}, David {Irusta}, Sidik {Isani}, Farbod {Jahandar},
  Manoj {Kaplinghat}, Collin {Kielty}, Neelesh {Kulkarni}, Leire {Lahidalga},
  Florence {Laurent}, Jon {Lawrence}, Mary~Beth {Laychak}, Jooyoung {Lee},
  Zhigang {Liu}, Nathan {Loewen}, Fernando {L{\'o}pez}, Thomas {Lorentz},
  Guillaume {Lorgeoux}, Billy {Mahoney}, Slavko {Mali}, Eric {Manuel},
  Sof{\'\i}a {Mart{\'\i}nez}, Celine {Mazoukh}, Youn{\`e}s {Messaddeq},
  Jean-Emmanuel {Migniau}, Shan {Mignot}, Stephanie {Monty}, Steeve {Morency},
  Yves {Mouser}, Ronny {Muller}, Rolf {Muller}, Gaizka {Murga}, Rick
  {Murowinski}, Victor {Nicolov}, Naveen {Pai}, Rafal {Pawluczyk}, John
  {Pazder}, Arlette {P{\'e}contal}, Andreea {Petric}, Francisco {Prada},
  Corinne {Rai}, Coba {Ricard}, Jennifer {Roberts}, J.~Michael {Rodgers}, Jane
  {Rodgers}, Federico {Ruan}, Tamatea {Russelo}, Derrick {Salmom}, Justo
  {S{\'a}nchez}, Will {Saunders}, Case {Scott}, Andy {Sheinis}, Douglas
  {Simons}, Scott {Smedley}, Zhen {Tang}, Jose {Teran}, Simon {Thibault},
  Sivarani {Thirupathi}, Laurence {Tresse}, Mitchell {Troy}, Rafael {Urrutia},
  Emile {van Vuuren}, Sudharshan {Venkatesan}, Kim {Venn}, Tom {Vermeulen}, Eva
  {Villaver}, Lew {Waller}, Lei {Wang}, Jianping {Wang}, Eric {Williams}, Matt
  {Wilson}, Kanoa {Withington}, Christophe {Y{\`e}che}, David {Yong}, Chao
  {Zhai}, Kai {Zhang}, Ross {Zhelem}, and Zengxiang {Zhou}.
\newblock {The Maunakea Spectroscopic Explorer Book 2018}.
\newblock {\em arXiv e-prints}, page arXiv:1810.08695, Oct 2018.

\bibitem{spectel}
Richard {Ellis} and Kyle {Dawson}.
\newblock {SpecTel: A 10-12 meter class Spectroscopic Survey Telescope}.
\newblock In {\em Bulletin of the American Astronomical Society}, volume~51,
  page~45, September 2019.

\bibitem{Dawson_Snowmass}
Kyle {Dawson}, Andrew {Hearin}, Katrin {Heitmann}, Mustapha {Ishak}, Johannes
  {Ulf Lange}, Martin {White}, and Rongpu {Zhou}.
\newblock {Snowmass2021 Cosmic Frontier White Paper: High Density Galaxy
  Clustering in the Regime of Cosmic Acceleration}.
\newblock {\em arXiv e-prints}, page arXiv:2203.07291, March 2022.

\bibitem{Ferraro_Snowmass}
Simone {Ferraro}, Noah {Sailer}, Anze {Slosar}, and Martin {White}.
\newblock {Snowmass2021 Cosmic Frontier White Paper: Cosmology and Fundamental
  Physics from the three-dimensional Large Scale Structure}.
\newblock {\em arXiv e-prints}, page arXiv:2203.07506, March 2022.

\bibitem{Bechtol_Snowmass}
Keith {Bechtol}, Simon {Birrer}, Francis-Yan {Cyr-Racine}, Katelin {Schutz},
  Susmita {Adhikari}, Arka {Banerjee}, Simeon {Bird}, Nikita {Blinov},
  Kimberly~K. {Boddy}, Celine {Boehm}, Kevin {Bundy}, Malte {Buschmann},
  Sukanya {Chakrabarti}, David {Curtin}, Liang {Dai}, Alex {Drlica-Wagner},
  Cora {Dvorkin}, Adrienne~L. {Erickcek}, Daniel {Gilman}, Saniya {Heeba},
  Stacy {Kim}, Vid {Ir{\v{s}}i{\v{c}}}, Alexie {Leauthaud}, Mark {Lovell},
  Zarija {Luki{\'c}}, Yao-Yuan {Mao}, Sidney {Mau}, Andrea {Mitridate}, Philip
  {Mocz}, Julian~B. {Mu{\~n}oz}, Ethan~O. {Nadler}, Annika H.~G. {Peter},
  Adrian {Price-Whelan}, Andrew {Robertson}, Nashwan {Sabti}, Neelima {Sehgal},
  Nora {Shipp}, Joshua~D. {Simon}, Rajeev {Singh}, Ken {Van Tilburg}, Risa~H.
  {Wechsler}, Axel {Widmark}, and Hai-Bo {Yu}.
\newblock {Snowmass2021 Cosmic Frontier White Paper: Dark Matter Physics from
  Halo Measurements}.
\newblock {\em arXiv e-prints}, page arXiv:2203.07354, March 2022.

\bibitem{Chakrabarti_Snowmass}
Sukanya {Chakrabarti}, Alex {Drlica-Wagner}, Ting~S. {Li}, Neelima {Sehgal},
  Joshua~D. {Simon}, Simon {Birrer}, Duncan~A. {Brown}, Rebecca {Bernstein},
  Alberto~D. {Bolatto}, Philip {Chang}, Kyle {Dawson}, Paul {Demorest}, Daniel
  {Grin}, David~L. {Kaplan}, Joseph {Lazio}, Jennifer {Marshall}, Eric~J.
  {Murphy}, Scott {Ransom}, Brant~E. {Robertson}, Rajeev {Singh}, An{\v{z}}e
  {Slosar}, Tommaso {Treu}, Yu-Dai {Tsai}, and Benjamin~F. {Williams}.
\newblock {Snowmass2021 Cosmic Frontier White Paper: Observational Facilities
  to Study Dark Matter}.
\newblock {\em arXiv e-prints}, page arXiv:2203.06200, March 2022.

\end{thebibliography}

\end{document}